\documentclass[aps,preprint,floatfix,pre]{revtex4-2}
\usepackage{graphicx} 
\usepackage{amssymb}
\usepackage{tikz}
\usepackage{amsmath}
\usepackage{booktabs}
\usepackage{multirow}
\usetikzlibrary{positioning,arrows.meta,calc, fit, backgrounds}
\usepackage{array}
\usepackage{makecell}
\usepackage{siunitx}
\newcommand{\w}{\boldsymbol{w}}
\newcommand{\p}{\boldsymbol{p}}

\begin{document}

\title{Sampling weights at intermediate temperatures is optimal for training large language protein models}
\author{L. Ghiringhelli}
\affiliation{Department of Physics, via Celoria 16, 20133 Milano, Italy}
\author{A. Zambon}
\email{alessandro.zambon@unimi.it}
\affiliation{Department of Physics and INFN, via Celoria 16, 20133 Milano, Italy}
\author{G. Tiana}
\email{guido.tiana@unimi.it}
\affiliation{Department of Physics and INFN, via Celoria 16, 20133 Milano, Italy}
\date{\today}

\begin{abstract}
    Using a statistical mechanics framework, we investigate the parameter space of transformer models trained on protein sequence data. We sample the loss landscape at varying temperatures using Langevin dynamics to characterize the low-loss manifold, and to understand the mechanisms underlying transformers' superior performance in protein structure prediction. We find that, at variance with networks not based on the attention mechanism, the lack of a first--order--like transition in the loss of the transformer produces a range of intermediate temperatures with good learning properties; this is true both for synthetic and natural protein sequences.  We also show that the parameters of most layers are highly conserved at these temperatures if the dimension of the embedding is optimal, and we provide an operative way to find this dimension. Additionally, we show that the attention matrix is more predictive of the contact maps of the protein at higher temperatures and for higher dimensions of the embedding than those optimal for learning. Finally, we showed that the models sampled at intermediate temperatures can predict the free-energy variation upon mutation, better than models obtained through standard optimization techniques.
\end{abstract}
\maketitle

\section{Introduction}

Large language models, based on the Transformer architecture \cite{Vaswani2023AttentionNeed}, have had a significant impact on a broad range of machine learning tasks. In particular, they have affected the field of biophysics, enabling computers to predict the native conformation of proteins from the knowledge of their sequences alone \cite{Jumper2021,Lin2022Evolutionary-scaleModel}. Despite their significance, the comprehension of their intrinsic mechanisms remains significantly less understood than that of other types of feedforward networks. 

In the case of simple feedforward networks, a large amount of information is now available about how they work, how they generalize from the training data and how the space of solutions looks like \cite{Seung1992StatisticalExamples,Monasson1994DomainsNetworks,Bahri2020,Baldassi2015,Baldassi2016UnreasonableSchemes,Rocks2022MemorizingModels,Becker2020GeometryNetworks,Zambon2025SamplingNetwork}. 

Apart from the fact that it is more cumbersome than training feedforward networks  with the same number of parameters but not based on the attention mechanism, less is known about the learning mechanism of transformers \cite{Liu2023UnderstandingTransformers}. In one--layer systems, gradient optimization was shown to act in two stages, first learning causal relations by amplifying attention weights between words that are distinctive in an example and suppressing those between common ones, and then refining the results logarithmically slowly \cite{Tian2023ScanTransformer,Nichani2024HowDescent}. 

Remarkably, it was proved analytically that a single transformer learns the couplings of a Potts model in a way that is equivalent to that obtained from pseudolikelihood learning \cite{Rende2024MappingModel}. Since Potts models describe well the coevolution of amino acids in proteins \cite{Morcos2011a}, transformers seem the ideal machine--learning tool to capture proximity relations in proteins.

As with the coevolution described by Potts models, transformers can be studied using the formalism of the canonical ensemble, thus treating them in the context of  statistical mechanics. This approach has been shown to be effective in the study of simple feedforward networks. In these networks, the low-loss manifold has been found to be fully connected and to have a spiky shape \cite{Zambon2025SamplingNetwork}.

In this work, we present results from sampling the parameters of a transformer to better understand how it works and why it efficiently retrieves spatial and chemical relationships between amino acids along the chain.
We compare these results with those from a pooled feedforward architecture lacking an attention mechanism. We framed the problem of exploring the parameter space in terms of statistical mechanics to leverage the computational algorithms developed in this field, in particular in connection with the canonical ensemble.

We sampled the space of parameters of the model with an algorithm based on the numerical solution of the Langevin equation, where the gradients are evaluated approximately on mini-batches of training sequences to gain computational efficiency \cite{Zambon2026ControlledMinibatches}. This algorithm relies on the control of the noise introduced by the approximation, adjusting in real time the masses and the friction coefficients of the Langevin equations to sample the canonical ensemble at the desired temperature. In this way, we combine the advantages of stochastic gradient descent, which is fast for optimization problems but does not samples specific probability distributions, and of Langevin equations, which samples Boltzmann's distribution but requires a lengthy calculation of the forces acting on the system.

Initially, we assigned a simple task to the network, that is learning a set of synthetic sequences that fold into a single structure. We chose this task because it is computationally efficient enough to collect statistics and draw robust conclusions about the system. This allows us to obtain information about how transformers function in general and their ability to predict protein structures. We then generalized our results, by studying the same problem for datasets composed of sequences belonging to different protein families.

In the next sections, we show that transformers display an intermediate range of temperatures with optimal generalization properties, while pooled feedforward networks have a sharper transition from low to high temperatures that prevents this regime (Sect. \ref{sect:temp_dep_prop} for the case of synthetic sequences). Moreover, we demonstrate that the parameters of most layers are highly conserved among models sampled at the optimal temperature, provided that the the embedding is properly dimensioned, and we suggest an operative way to find this ideal dimension (Sect. \ref{sect:var_par}). Additionally, we show that the attention matrix is more predictive of the contact maps of the protein at temperatures and for dimensions of the embedding that are higher than those optimal for learning (Sect. \ref{sect:att_mat}). Finally, we obtain the same qualitative thermodynamic results for real sequence data and we demonstrate that, on average, sampling at intermediate temperatures can improve the generalization performance and the prediction of free-energy variation due to single-site mutations with respect to standard optimization techniques (Sect. \ref{sect:Pfam}).

\section{Datasets, models and algorithms}
\label{sect:models_data_algo}

    \subsection{The synthetic datasets}
    \label{sect:synthetic_dataset}

        \begin{figure}
            \centering
            \includegraphics[width=\linewidth]{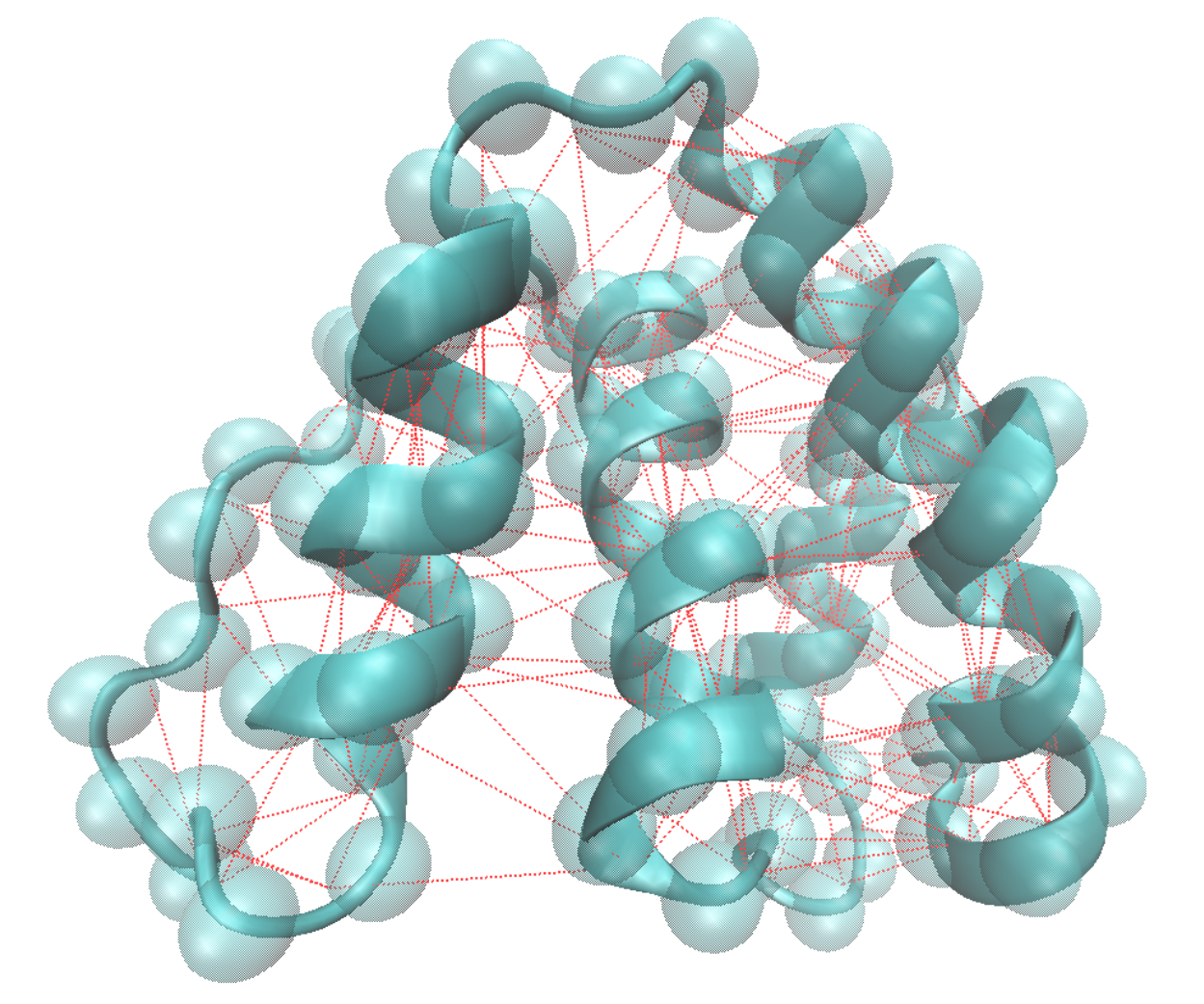}
            \caption{The structure of acyl--coenzyme A binding protein (pdb code: 2abd), where the contacts between amino acids are indicated with dashed red lines.}
            \label{fig:2abd}
        \end{figure}
        The task we addressed is a masked language modeling problem, for a synthetic dataset of amino acid sequences folding to the native structure of the acyl--coenzyme A binding protein (Fig. \ref{fig:2abd}, pdb code: 2abd) \cite{Kragelund1995}.
        
        The dataset was generated following the same strategy as in ref. \cite{Franco2019}: we defined the energy of a sequence $\boldsymbol{x}$ as
        \begin{equation}
            U(\boldsymbol{\boldsymbol{x}})=\sum_{ij}B_{x_ix_j}\Delta_{ij},
        \end{equation}
        where $B_{xy}$ is a randomly--generated symmetric interaction matrix between amino acids, extracted from a Gaussian distribution with zero average and unitary variance, while $\Delta_{ij}$ is a binary symmetric contact matrix, derived from the native conformation of the acyl--coenzyme A binding protein so that $\Delta_{ij} = 1$ if the corresponding amino acids $x_i$ and $x_j$ have a pair of heavy atoms closer than $3.5$\AA, a standard threshold that accounts for atomic interactions.
        In the canonical ensemble framework, we generated $15000$ low--temperature ($T_s=1.27$, in energy units; cf. ref. \cite{Franco2019}) sequences $\boldsymbol{x}$ of length $n=86$ with an adaptive simulated tempering algorithm \cite{Tiana2011}.

        For each generated sequence $\boldsymbol{x}^\mu$, we selected approximately $15\%$ of its amino acids and replaced them with a masking token. The true amino acids in the masked sites were gathered in a vector $\boldsymbol{y}^\mu$, which was kept fixed for each sequence.
        Then, we partitioned the whole dataset $\{\left( \boldsymbol{x}^\mu, \boldsymbol{y}^\mu \right)\}_{\mu=1}^{15000}$ into three subsets: train, validation and test, composed of $1000$, $1000$ and $13000$ pairs respectively.

    \subsection{The Pfam datasets}
    \label{sect:Pfam_dataset}
        We addressed the same masked language modeling problem described in Sect. \ref{sect:synthetic_dataset}, using sequences belonging to 10 different protein families, as defined in the Pfam database \cite{Punta2012}. Each family was associated with a different dataset, where, as previously, we masked approximately $15\%$ of the amino acids of each sequence and replaced them with a masking token. Each dataset was then partitioned into three subsets: train, validation and test, corresponding to $70\%$, $15\%$ and $15\%$ of the total, respectively.
        The generation process of each Pfam dataset, along with the list of the selected protein families, can be found in Appendix \ref{app_sect:Pfam_dataset}.

    \subsection{The prediction models}
    \label{sect:models}

        \begin{figure}
            \centering
            \includegraphics[width=\linewidth]{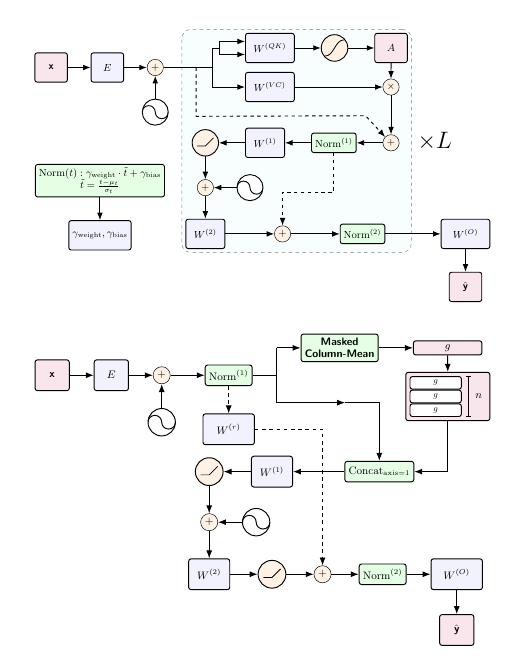}
        \caption{
        The scheme of the $L$-module transformer (upper scheme) and of the pooled feedforward model (lower scheme) used in this work.
        }
        \label{fig:scheme}
        \end{figure}

        We selected three different architectures to tackle the masked language modeling problems described in Sect. \ref{sect:synthetic_dataset} and in Sect. \ref{sect:Pfam_dataset}. Two of these are transformers \cite{Vaswani2023AttentionNeed}, made up of either one ($\mathrm{TF}_1$) or four ($\mathrm{TF}_4$) modules in series (upper panel in Fig. \ref{fig:scheme}), and the third is a pooled feedforward network ($\mathrm{FF}$) whose number of parameters ($1.74\times 10^5$) is approximately equal to that of the 4-module transformer ($1.69\times 10^5$), and four times that of TF$_1$ ($4.4\times 10^4$ parameters).     
        
        All of these models follow a three--stage structure comprising an input construction stage, an encoder and a final output projection layer. The input and output components are the same for the two network families studied here, but the internal encoder differs between transformer-based and pooled feedforward networks.
        
        The encoders of transformer-based models use the attention mechanism to efficiently capture long-range dependencies in the input and produce a context-aware representation of each input component.
        
         In the pooled feedforward encoder, we modeled the architecture to resemble a transformer as closely as possible, replacing only the attention matrix with a pooling layer. This allows us to isolate the effects of the attention mechanism.
        Furthermore, each architecture is designed so that all possible permutation symmetries are eliminated without impairing the learning capabilities of the model.
        The detailed description of  each neural network can be found in Appendix \ref{app_sect:models}.

        For a given masked input, each model produces a matrix. Every line of the matrix represents the predicted probability distribution over the vocabulary tokens (in this case, amino acid types) for the corresponding site along the sequence. The quality of this prediction has been evaluated using the cross--entropy loss function and the error function (see Appendix \ref{app_sect:observables} for mathematical details).

    \subsection{The sampling algorithm}
    \label{sect:algo}

        The space of parameters of the system is sampled according to Boltzmann's probability, using the loss as the energy of the system and considering it in the canonical ensemble of statistical mechanics. In fact, Boltzmann's distribution is the maximum-entropy distribution at fixed average loss; sampling the parameter space at different temperatures $T$ allowed us to study the manifold corresponding to different loss values without biasing the exploration. Moreover, we could take advantage of the large set of theoretical and computational tools developed in the realm of statistical mechanics.  

        Operatively, we first minimized the loss of each network with the Adam  optimizer (cf. App.  \ref{app_sect:training}), producing two types of trained weight vectors, namely early-stopped (ES) and over-trained (OT). 
       Then, we sampled the space of all the parameters that define the networks using a norm--constrained version (cf. Appendix \ref{app_sect:sampling}) of the pseudo--Langevin sampling algorithm, recently developed for high--dimensional parameter spaces like those of artificial neural networks \cite{Zambon2026ControlledMinibatches}. It basically consists of solving the Langevin equations, using approximate forces evaluated on mini--batches and adjusting the fictitious masses associated with the parameters in real time to guarantee that the noise term of the dynamical equations is thermalized at the correct temperature. 

       At each temperature, we could thus collect a large number of sets of parameters, corresponding to different prediction models. We used these sets to estimate average quantities over the sampled models (such as the average validation error) and probability distributions (such as the similarity distribution between the parameters of the models). Assuming the system is ergodic, the averages converge to the thermodynamic averages and the probability distributions to Boltzmann's.

\section{Results}

    We studied the properties of models sampled according to Boltzmann distribution at different temperatures when trained either with synthetic sequences (Sects. III A, B, C) or with families of natural sequences (Sect. III D).

    \subsection{The generalization error is minimal at intermediate temperatures}
    \label{sect:temp_dep_prop}

\begin{figure}
            \centering
            \includegraphics[width=\linewidth]{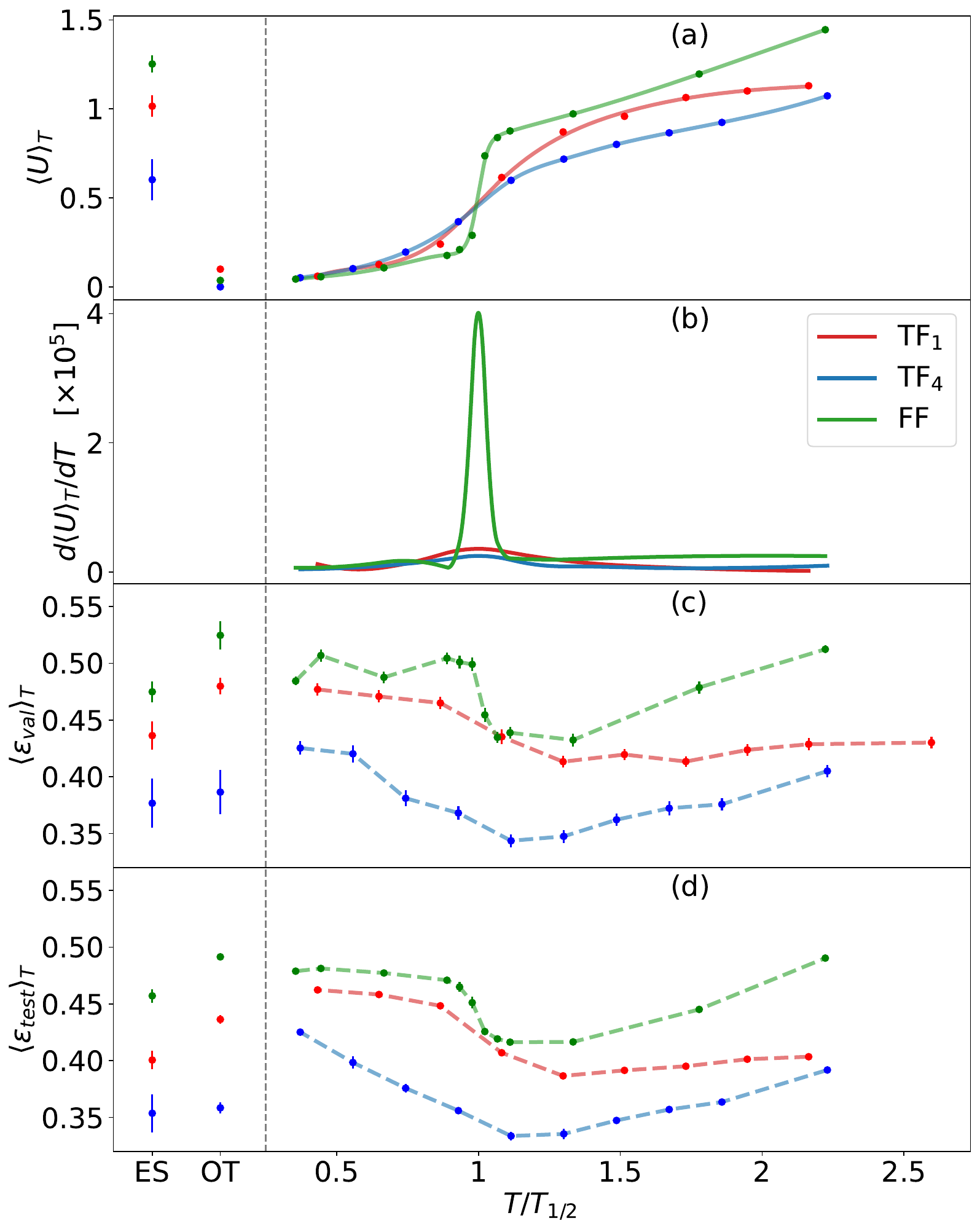}
            \caption{
            Right of the line, the average value of the potential energy $\langle U \rangle_T$ (a), of the specific heat $d\langle U\rangle_T/dT$ (b), of the validation error $\langle \epsilon_{\mathrm{val}} \rangle_T$ (c), and of the test error $\langle \epsilon_{\mathrm{test}} \rangle_T$ (d) as functions of the rescaled temperature $T/T_{1/2}$ for the models $\mathrm{TF}_1$  (red curves), $\mathrm{TF}_4$  (blue curves) and $\mathrm{FF}$ (green curves).
            Left of the line, the same for early–stopped solutions (ES) and over-trained solutions (OT) obtained with Adam.
            Circles denote numerical measurements, while solid lines represent spline interpolations. All points have an error bar which indicates the standard deviation.
            }
            \label{fig:termodinamica}
        \end{figure}

        We compared the sampling of the parameter space at different temperatures for the single--block transformer TF$_1$, the four-block transformer TF$_4$ and the pooled feedforward network FF, acting on the same data with the same task (cf. Sect. \ref{sect:models}). The comparison is extended to the results obtained with the standard optimization of the parameters by gradient--descent techniques, that in the present context can be regarded as zero--temperature calculations. For each architecture type, the system is in the overparametrized regime, as clear from the fact that OT solutions and low--temperature weight vectors are associated to a training error equal or close to 0 (Fig. \ref{app_fig:termodinamica_tot}b and Table \ref{app_tab:train_insight} in the appendices). The most apparent difference between transformers and pooled feedforward architectures is in the behavior of the average loss as a function of the temperature (Fig. \ref{fig:termodinamica}a). While the pooled feedforward network displays a sharp jump in the loss (corresponding to a sharp peak in the specific heat, Fig. \ref{fig:termodinamica}b), reminiscent of a first--order phase transition, both transformers display a much smoother shape. For the purposes of comparison, we have normalized the temperature with respect to the temperature $T_{1/2}$ at the maximum in the specific heat (which is architecture--dependent), in order to make the temperature scales of the different models comparable.
        
        The accuracy of the prediction on the test set quantifies the ability of the model to generalize. The generalization accuracy for all models has a non-monotonic shape, reaching a maximum right after the transition temperature (Fig. \ref{fig:termodinamica}d). This intermediate--temperature regime can be easily identified on the fly, looking at the error function computed on the validation set (Fig. \ref{fig:termodinamica}c). In this range of  temperatures, the four--block transformer has the greatest generalization ability. In any case, the one--block transformer performs better than the feed-forward architecture, despite the size difference between the two models (cf. Table \ref{app_tab:model_configs} in the appendices).

       The OT weigths of $\mathrm{TF}_{1}$ and $\mathrm{FF}$ display loss, training and generalization errors similar to the averages at low temperatures (Fig. \ref{fig:termodinamica}), in agreement with the expectation that OT corresponds to zero temperature. On the contrary, the OT weights of $\mathrm{TF}_{4}$, has much lower generalization error, while the loss is again what expected at zero temperature. The ES weights show an average loss that is similar to that obtained from sampling at intermediate temperatures, and they consistently demonstrate higher generalization ability. However, the average test error $\epsilon_{test}$ over the ES vectors is always larger than the average at the best temperature (by 0.01 at $T^{\mathrm{best}} =1.3\,T_{1/2}$ for $\mathrm{TF}_1$, by 0.02 at $T^{\mathrm{best}} =1.1\,T_{1/2}$ for $\mathrm{TF}_4$ and by 0.04 at $T^{\mathrm{best}} =1.1\,T_{1/2}$ for $\mathrm{FF}$). Moreover, the variability of the generalization error for different realizations of the ES minimization is much larger than that observed in the sampling at intermediate temperature (error bars in Fig. \ref{fig:termodinamica}d).
       
    \subsection{Varying conservation of the parameters in the different layers}
    \label{sect:var_par}

        We then examined the variability of the parameters in the different layers of the networks during the sampling. For this purpose, we calculated the distribution of the mutual similarity 
        \begin{equation}
            q^l_{t,t'}= \frac{\w^{l}(t)\cdot\w^{l}(t')}{|\w^{l}(t)||\w^{l}(t')|}
        \end{equation}
        of each layer $l$ over the sampling, where $\w^{l}(t)$ is the vector containing the parameters of the layer $l$ sampled at time $t$ during a simulation at temperature $T^{\mathrm{best}}$ (see Sect. \ref{sect:temp_dep_prop}).
        We could compare the weights element--wise because the permutation symmetries of the networks had been broken by construction (see Appendix \ref{app_sect:models}).
        
\begin{figure}
            \centering
            \includegraphics[width=\linewidth]{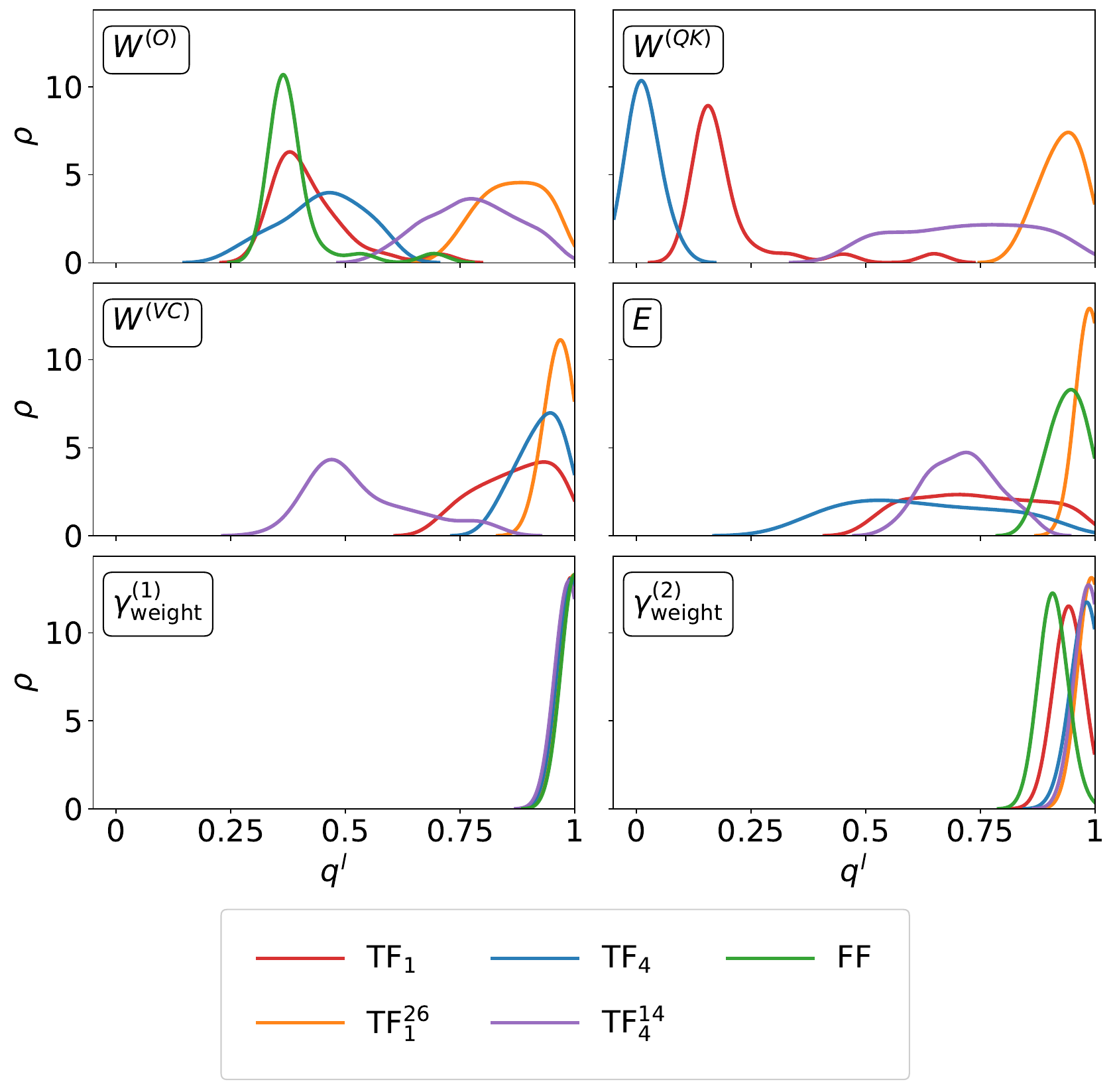}
            \caption{
            The distributions $\rho$ of the similarity $q^l$ of parameters for some of the layers, obtained by comparing equilibrated configurations at temperature $T^{\mathrm{best}}$ for each architecture (identified by different colors). The complete set is displayed in Fig. \protect\ref{app_fig:q_tot} in the Appendix. The architectures labeled as $\mathrm{TF}_1^{26}$ and $\mathrm{TF}_4^{14}$ correspond to transformers of reduced dimensionality (see text).
            }
            \label{fig:q}
        \end{figure}

        In the one--block transformer, the distribution $\rho(q^l)$ of mutual similarity between pairs of models is either peaked around zero or very broad for most layers, indicating that the network can find several sets of parameters that maintain a sufficiently low loss (Fig. \ref{app_fig:q_tot} in the appendices). There are, however, exceptions. The parameters $\boldsymbol{\gamma}^{(1)}_\mathrm{weight}$ and $\boldsymbol{\gamma}^{(2)}_\mathrm{weight}$ that weight the rows in the normalization layers display similarity distribution strongly peaked towards 1, indicating that the values are well conserved throughout the sampling (red curves in the lower panels in Fig. \ref{fig:q}). The parameters $W^{(VC)}$ which control the value layer and $E$ controlling the embedding are also conserved, although their distribution is more broadened than that of the normalization layers (Fig. \ref{fig:q}, red curves in the middle panels). The parameters $W^{(O)}$ controlling the output and the query--key parameters $W^{(QK)}$ are peaked at intermediate values (Fig. \ref{fig:q}, upper panels, red curves). Different layers of $\mathrm{TF}_1$ are then conserved at varying degree at the optimal temperature.
        
        In the case of a network made of 4 transformer blocks, the variability of the parameters in each block is very similar to that of the $\mathrm{TF}_1$ system (the distributions for the first layer are shown as blue curves in Fig. \ref{fig:q}). An exception is the set of value parameters $W^{(VC)}$ that is peaked around 0 in the last three blocks (Fig. \ref{app_fig:q_tot}), while it is peaked around 1, as for the single-module architecture, only in the first block of the transformer.
        
        The distribution of similarities in the layers of the pooled feedforward network $\mathrm{FF}$ (Fig. \ref{fig:q}, green curves) is overall quite similar to that of the corresponding layers in the transformers, suggesting that their conservation are not a consequence of the attention mechanism. 
        
\begin{figure*}
            \centering
            \includegraphics[width=0.8\linewidth]{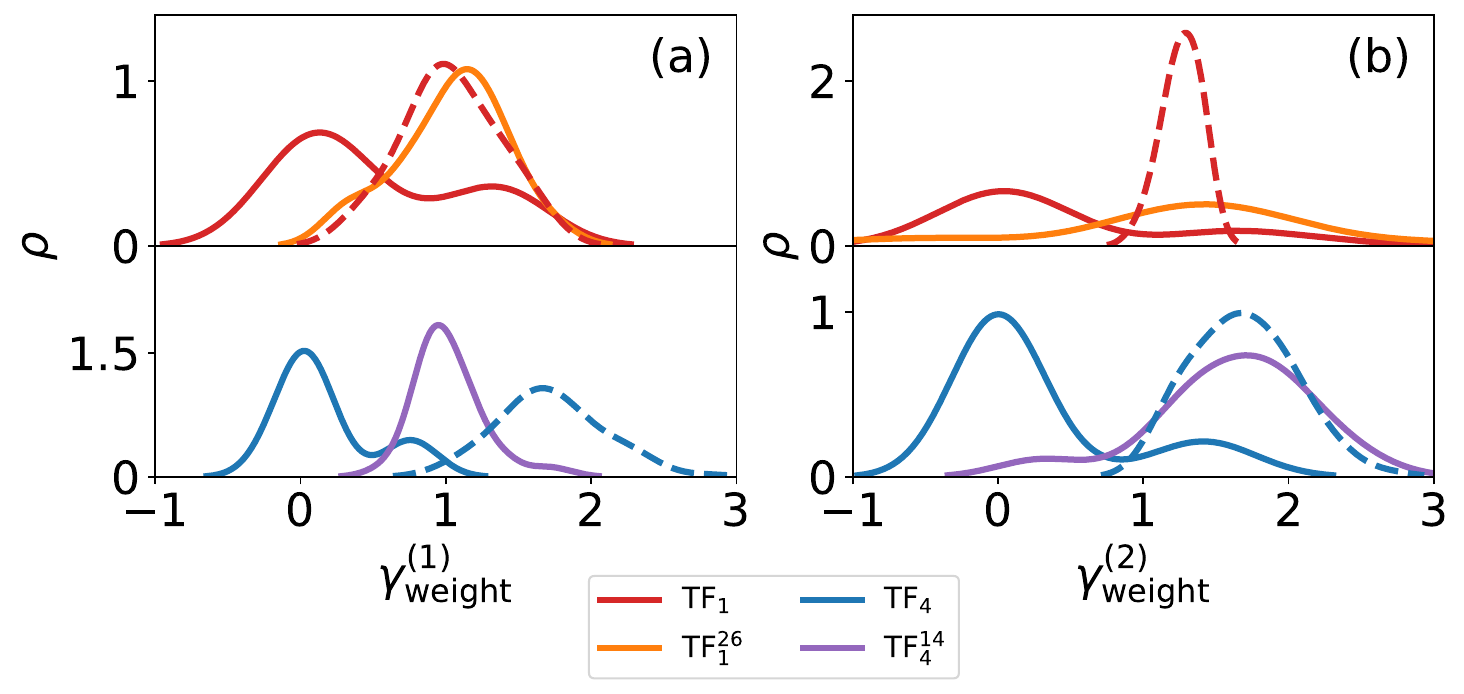}
            \caption{
            The distribution $\rho$ of the values of $\boldsymbol{\gamma}_{\mathrm{weight}}^{(1)}$ (a) and $\boldsymbol{\gamma}_{\mathrm{weight}}^{(2)}$ (b) obtained from the sampling at $T^{\mathrm{best}}$ of the original and reduced transformers (solid lines).
            The dotted red lines indicate the distribution of $\boldsymbol{\gamma}_{\mathrm{weight}}^{(1/2)}$ values obtained from the ES minimization.
            The values reported for the four--module transformers refer to the last module.
            }
            \label{fig:gamma}
        \end{figure*}

        The strong conservation of the weights $\boldsymbol{\gamma}_{\mathrm{weight}}^{(1)}$ and $\boldsymbol{\gamma}_{\mathrm{weight}}^{(2)}$ in all architectures facilitated the inspection of the values they assume at during the sampling. For the $\mathrm{TF}_1$ model, the distribution is bimodal (Fig. \ref{fig:gamma}, solid red curves), with one peak centered around zero (amounting to approximately $60\%$ of the elements for $\boldsymbol{\gamma}_{\mathrm{weight}}^{(1)}$ and $80\%$ of the elements for $\boldsymbol{\gamma}_{\mathrm{weight}}^{(2)}$) and the other one around a positive value. Since these vectors have size equal to the embedding dimension $d=64$ and they weight the information after the attention matrix, we hypothesized that the elements with value close to zero are essentially useless for the network's task. 


        Interestingly, the distribution of both sets of parameters has a single-peaked distribution for ES weights, centered at non-zero values. This is similar to the higher peak obtained through sampling for $\mathrm{TF}_{1}$ (Fig. \ref{fig:gamma}, dashed red curves), which suggests that our hypothesis would not apply to the parameters obtained from loss minimization.
        
        To verify the hypothesis, we trained a new network (details in Appendix \ref{app_sect:training}), where the dimension of the embedding is reduced to $d=26$, corresponding to the elimination of the rows associated with values of $\boldsymbol{\gamma}_{\mathrm{weight}}^{(1)}$ close to zero. This new architecture will be referred to as $\mathrm{TF}_1^{26}$ (cf. Table \ref{app_tab:model_configs} in the appendices). We performed a sampling of the parameters at temperature $T^{\mathrm{best}} \approx  1.5 \, T_{1/2}$ obtaining an average generalization error of $\langle \epsilon_\mathrm{test}\rangle_{T^\mathrm{best}} = 0.40 \pm 0.001$, which is comparable to that of the original larger model $\langle \epsilon_\mathrm{test}\rangle_{T^\mathrm{best}} = 0.38\pm0.002$.
        As expected, the values assumed by the new weight vectors $\boldsymbol{\gamma}_{\mathrm{weight}}^{(1/2)}$ during the sampling are now distributed according to a single-peaked distribution at non-zero values (Fig. \ref{fig:gamma}, orange curves) and remain very conserved during the sampling (Fig. \ref{fig:q}, orange curves in the lower panels).
        
        Moreover, the per--layer similarities of the matrices $E$, $W^{(VC)}$, $W^{(QK)}$ and $W^{(O)}$ for the compact architecture $\mathrm{TF}_1^{26}$ are now all peaked toward higher values (Fig. \ref{fig:q}, orange curves). This suggests that the high variability these layers had in the larger system was primarily due to their oversized dimensions. Summing up, in a single--block transformer whose embedding dimension is carefully tuned, only the layers $W^{(1)}$, $W^{(2)}$ after the attention and the biases $b^{(0)}$ of the output can vary quite freely along the sampling, while the other layers are essentially fixed.

        This line of reasoning can also be applied to the $\mathrm{TF}_{4}$ architecture. We examined the distributions of the values assumed by the vectors $\boldsymbol{\gamma}_{\mathrm{weight}}^{(1/2)}$ in the last transformer block, observing that they have qualitatively the same distribution we identified in the single--block model (Fig. \ref{fig:gamma}, blue curves).  Again, we trained a new network with a reduced embedding dimension $d=14$ (called $\mathrm{TF}_4^{14}$), corresponding to the number of non--zero values of $\boldsymbol{\gamma}_{\mathrm{weight}}^{(1)}$ in the last block. The sampling at $T^{\mathrm{best}} \approx  1.7 \, T_{1/2}$ of the parameters of this reduced architecture show a distributions $\rho \left( \boldsymbol{\gamma}_{\mathrm{weight}}^{(1/2)} \right)$ which is unimodal and centered at non--zero values (Fig. \ref{fig:gamma}, purple curves). The average generalization error is $\langle \epsilon_\mathrm{test}\rangle_{T^\mathrm{best}} = 0.36 \pm 0.002$, comparable to that of the $\mathrm{TF}_4$ (i.e. $\langle \epsilon_{\mathrm{test}}\rangle_{T^\mathrm{best}} = 0.33 \pm 0.004$). The similar generalization ability occurs in spite of the huge difference in size between the two models (cf. Table \ref{app_tab:model_configs} in the appendices). For the different layers of $\mathrm{TF}_4^{14}$ the similarity of the sampled parameters is overall smaller than for the one--block architecture, but is anyhow larger, in general, than $0.5$, with the exception of biases which can vary much from model to model (Fig. \ref{fig:q}, purple curves).
        
    \subsection{The attention matrix is optimal at higher temperatures and larger embeddings}
    \label{sect:att_mat}
        
        When applied to the structure prediction of proteins, a key property of transformers is that their attention matrix resembles the adjacency matrix of the input sequence \cite{Rao2020TransformerLearners}.
        
        For the one--block transformer $\mathrm{TF}_1$, the mean attention matrix (i.e., the attention matrix symmetrized and averaged over the test set sequences) for models sampled at intermediate temperatures ($T=2T_{1/2}$) resembles the experimental contact map (cf. upper panels in Fig. \ref{fig:attention} and Fig. \ref{app_fig:contact_map} in the appendices). However, the same is not true for the mean matrices associated with ES or OT vectors (cf. Fig. \ref{app_fig:attention_app}a and b, respectively). 
        The mean attention matrix can be regarded as a self--averaging quantity, since individual models sampled at the same temperature produce mean matrices which are very similar to each other (e.g., Fig. \ref{app_fig:attention_app}d, f, g in the appendices).

        The mean attention matrix of one of the blocks of the $\mathrm{TF}_4$ model has a better similarity to the experimental contact map (lower panels in Fig. \ref{fig:attention}) than that of $\mathrm{TF}_1$. In this example, the map giving the improved agreement is that belonging to the third attention block of a model sampled at $T=1.1\,T_{1/2}$. The mean attention matrix can capture most contacts within alpha--helices and many tertiary contacts. Most false--positive contacts are actually between residues that are adjacent to residues forming true contacts. 
        
        The mean attention matrices belonging to the other blocks are very dissimilar from the contact map of the protein (cf. Fig. \Ref{app_fig:attention_app}j, k, l in the appendices). In another independent sampling we conducted, the mean matrix that resembled the experimental contact map ranged from the third to the fourth. The mean attention matrix of the best--matching block seems self--averaging also in the $\mathrm{TF}_4$ model (e.g., Fig. \Ref{app_fig:attention_app}m, n, o).

        At variance with the $\mathrm{TF}_1$ model, in the $\mathrm{TF}_4$ also ES and OT give mean matrices reasonably predictive of the contact map, though they perform worse than those obtained through sampling (cf. with Fig. \Ref{app_fig:attention_app}g, h with Fig. \Ref{app_fig:attention_app}m, n, o).
        
        A metric commonly used to assess the model capability of predicting protein contacts is the 'precision at L' (P@L) \cite{Rao2021MSATransformer}, which is the fraction of correct contacts predicted from the highest $L$ elements of the attention matrix. 
        In the present case, the metric is computed on the mean attention matrices, with $L=86$ (i.e. the length of the protein sequence). The fraction of correctly predicted contacts in the case of very large language models ranges typically between 0.4 and 0.6 \cite{Vig2021BERTologyModels}.

\begin{figure}
            \centering
            \includegraphics[width=\linewidth]{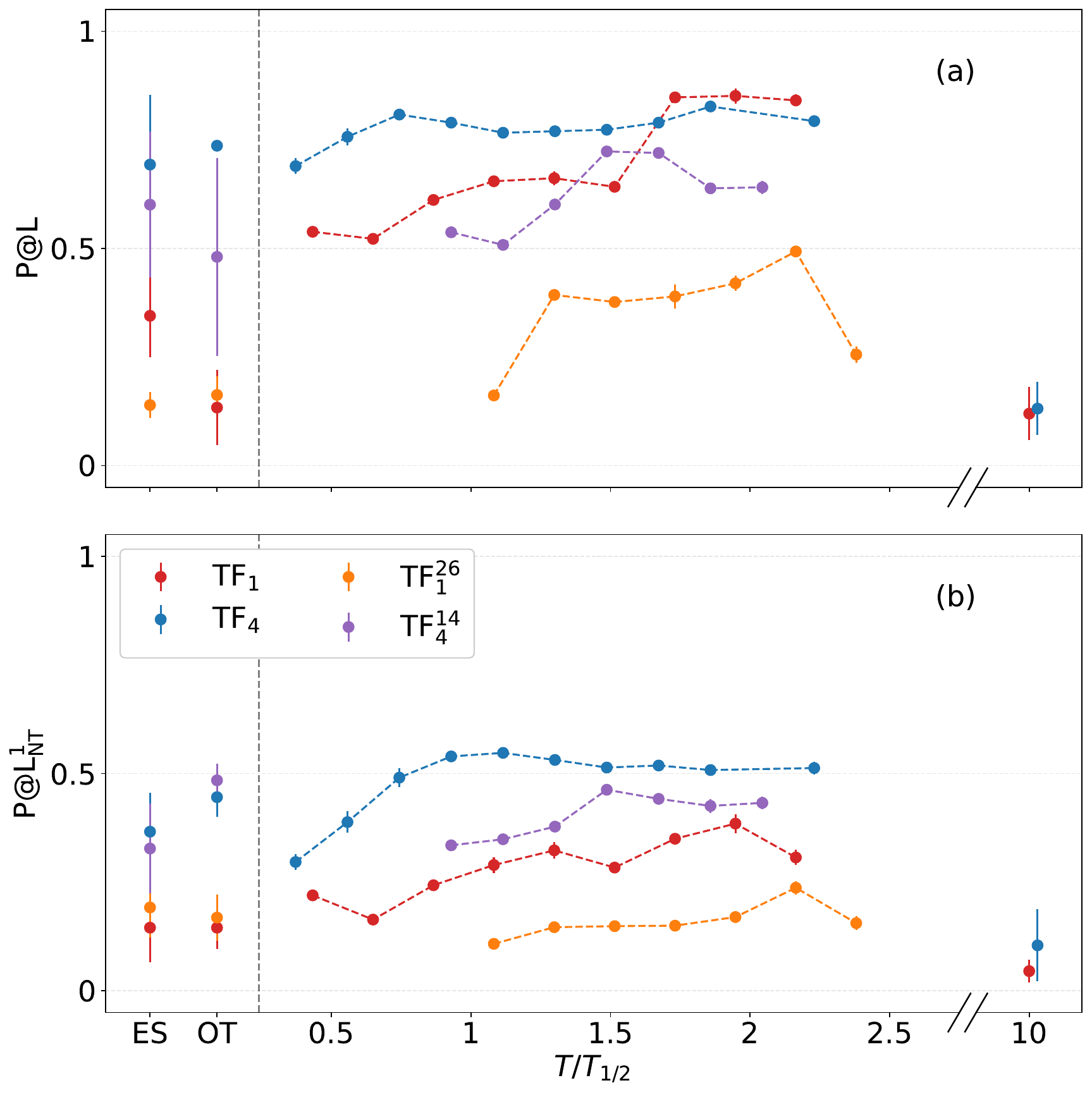}
            \caption{
            The average and standard deviation of the $\mathrm{P@L}$ (a) and of the $\mathrm{P@L}_{\mathrm{NT}}^{1}$ (b) metrics for ES and OT vectors (left of the line) and for models obtained from sampling at different rescaled temperatures $T/T_{1/2}$, for each architecture type. 
            }
            \label{fig:P@L}
        \end{figure}

        We calculated the P@L metric by averaging over the models sampled at each temperature for both $\mathrm{TF}_1$ and the best--matching block of $\mathrm{TF}_4$. We repeated the same calculation for the ES and OT solutions (Fig. \ref{fig:P@L}a). However, we observe that this metric does not recapitulate the intuition that $\mathrm{TF}_4$ is more predictive than $\mathrm{TF}_1$, as suggested by Fig. \ref{fig:attention}. In fact, the P@L prioritizes, in the list of the $L$ top scores, contacts close to the diagonal, not providing a metric that quantifies the ability of the mean attention matrices to predict the contact map as a whole.
       
        To solve this problem, we introduce a new metric, $\mathrm{P@L}_\mathrm{NT}^\Delta$, which takes into account the contacts between residues $i$ and $j$ only if $|j-i|>7$. Additionally, it defines a predicted contact $\left( i_\mathrm{p}, j_\mathrm{p} \right)$ as correct if it is within an interval $\Delta$ from a true contact $\left( i_\mathrm{t}, j_\mathrm{t} \right)$  (i.e., if $|i_\mathrm{p} - i_\mathrm{t}| + |j_\mathrm{p}-j_\mathrm{t}| \leq \Delta$). According to the new metric, $\mathrm{TF}_4$ is more predictive at all intermediate temperatures than $\mathrm{TF}_1$ (Fig. \ref{fig:P@L}b). The overall lower values of $\mathrm{P@L}_\mathrm{NT}^1$ with respect to $\mathrm{P@L}$ are due to the the fact that the former is agnostic of the alpha--helices, which are anyway easy to predict with other algorithms \cite{McGuffin2000}. Increasing the value of $\Delta$ in the definition of $\mathrm{P@L}_\mathrm{NT}^\Delta$  does not change qualitatively the results, just shifting the curves upwards (cf. Fig. \ref{app_fig:P@L_tot}).

        For the $\mathrm{TF}_1$ and $\mathrm{TF}_4$ architectures, the mean attention matrices obtained by sampling perform much better than those obtained from training (Fig. \ref{fig:P@L}b, blue and red points). This improvement is smaller (for $\mathrm{TF}_1^{26}$) or absent (for $\mathrm{TF}_4^{14}$) for transformer with reduced embedding dimension.
     
        Interestingly, there is not a straightforward correlation between generalization error and contacts prediction capability (cf. Fig. \ref{fig:termodinamica}d and Fig. \ref{fig:P@L}b). The values of $\mathrm{P@L}_\mathrm{NT}^1$ for the $\mathrm{TF}_{4}$ model stay constantly high in the whole range of intermediate temperatures, whereas $\langle\epsilon_\mathrm{test}\rangle_T$ increases shortly after $T_{1/2}$. For the $\mathrm{TF}_{1}$ architecture the best--matching mean attention matrices are found at temperatures consistently higher than $T_{1/2}$. We then conclude that the temperatures which provide predictive attention matrices do not match those at which the prediction of the protein sequence is optimal.
            
        The models with a reduced embedding dimension $d$ (i.e. $\mathrm{TF}_1^{26}$ and $\mathrm{TF}_4^{14}$) exhibit a similar behavior. However, their performance is clearly worse (Fig. \ref{fig:P@L}b). Thus, while these minimal models can reproduce protein sequences as well as those with a larger embedding dimension, their ability to predict contact maps is much poorer. This further points to the fact that contact prediction (for which the network has not been trained) is not completely correlated with sequence prediction. Modifying the models to improve the training and the generalization error thus is not a good strategy to improve the predictivity of the attention matrix. 
        This is also confirmed by the better performance of the OT models with respect to the ES solutions for almost every architecture (and especially for $\mathrm{TF}_{4}$ and $\mathrm{TF}_{4}^{14}$, Fig. \ref{fig:P@L}b).
        
        \begin{figure*}
            \centering
            \includegraphics[width=\linewidth]{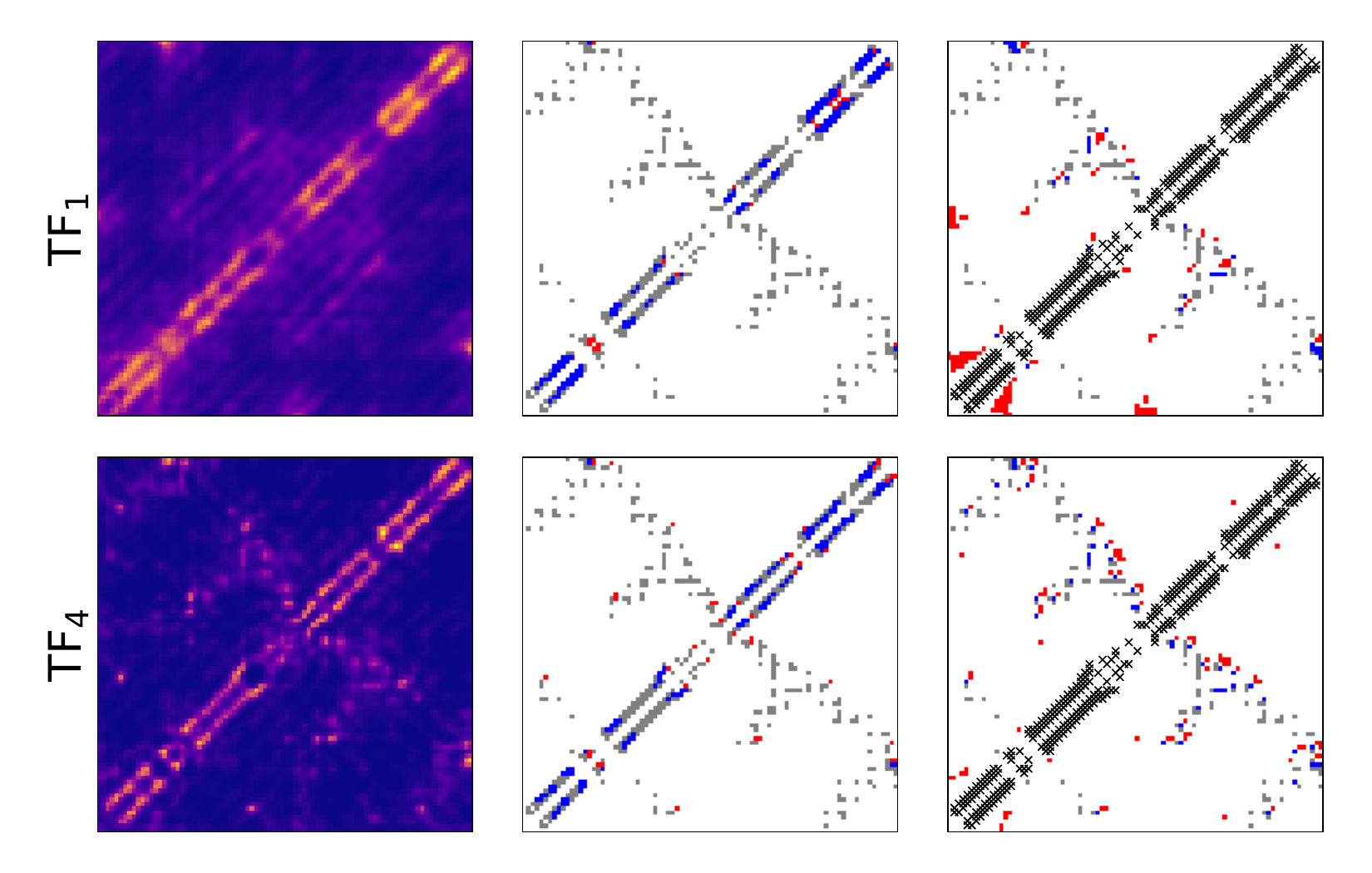}
            \caption{
            Left column, the mean attention matrix, symmetrized and averaged over the test set sequences, for the $\mathrm{TF}_{1}$ (upper row) and for the $\mathrm{TF}_{4}$ (lower row) models, obtained during a sampling a temperature $T = 2\,T_{1/2}$ and $T=1.1 \, T_{1/2}$ respectively. 
            Center column, the contact map obtained from the mean attention matrix using $\mathrm{P@L}$; true positives are in blue, false positives in pink and false negatives in gray. Right column, the contact map obtained from the mean attention matrix using $\mathrm{P@L}_{\mathrm{NT}}^{1}$. 
            }
            \label{fig:attention}
        \end{figure*}
        
    \subsection{Application to natural proteins}
    \label{sect:Pfam}

\begin{figure*}
            \centering
            \includegraphics[width=\linewidth]{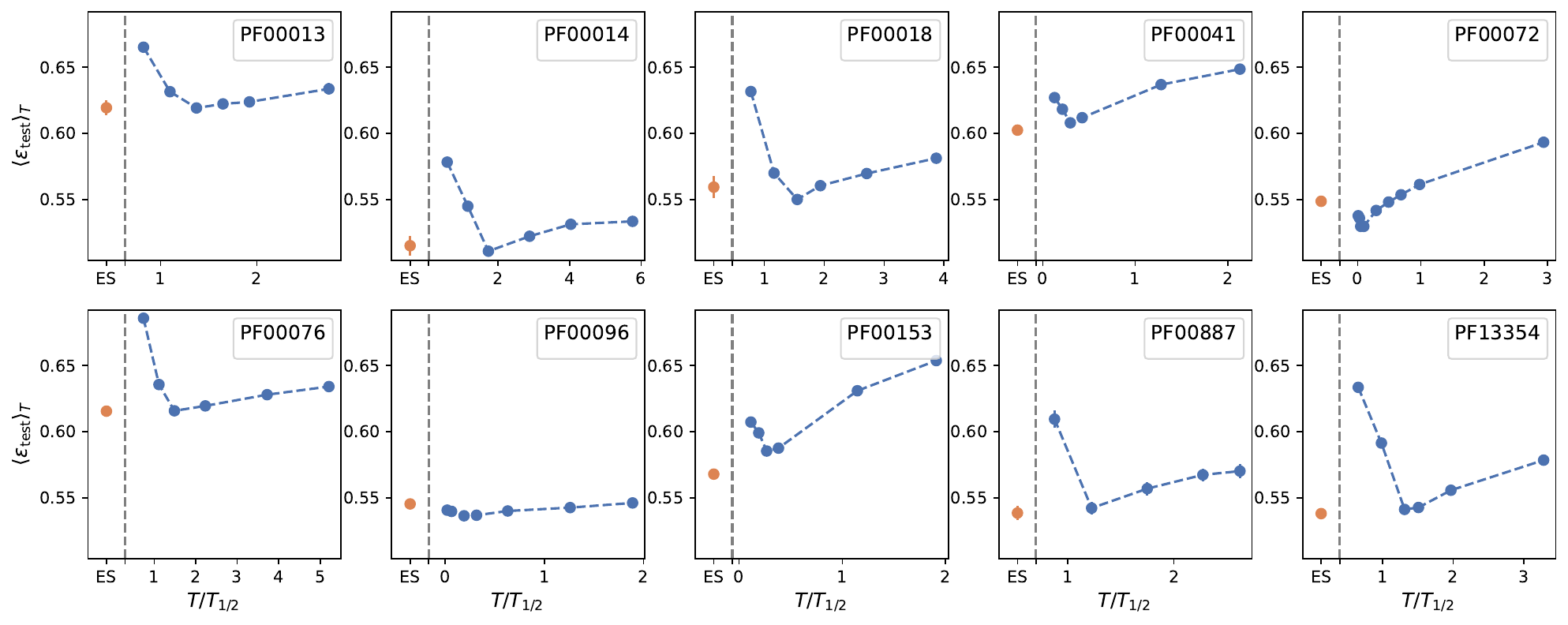}
            \caption{
            For each panel, at the right of the gray dotted line, the average value of the test error $\langle \epsilon_{\mathrm{test}} \rangle_T$ as function of the rescaled temperature $T/T_{1/2}$ for the $\mathrm{TF}_4$ model.
            At the left of the line, the same for early–stopped solutions (ES) obtained with Adam.
            All points have an error bar which indicates the standard deviation.
            }
            \label{fig:Pfam_thermodynamics}
        \end{figure*}

       We then studied 10 families of natural proteins, to validate the results obtained for the synthetic sequences (Sect. \ref{sect:att_mat}) in the real--life case, where evolution is driven by constraints that are more complex than thermodynamic stability. For each protein family (see Sect. \ref{sect:Pfam_dataset}), we sampled the space of parameters of the four-block transformer $\mathrm{TF}_4$ at different temperatures (Fig. \ref{fig:Pfam_thermodynamics}, blue dots). As for the synthetic scenario, we observed a range of intermediate temperatures  where the test error has a minimum. Moreover, in most cases the optimal temperature $T^\mathrm{best}$ is found just after the peak in specific heat ($T/T_{1/2} \approx 1$), as for synthetic proteins. There are some exceptions, namely protein families PF$00041$, PF$00096$ and PF$00153$, where the best generalization performance is obtained at lower temperatures ($T/T_{1/2} \ll 1$).
        
       We also compared the sampled averages with the results obtained with standard optimization of the parameters by gradient--descent techniques (Fig. \ref{fig:Pfam_thermodynamics}, orange dots), as  already done in Sect. \ref{sect:temp_dep_prop} for synthetic sequences. As a rule, the generalization capabilities of the network are either improved (as for PF$00018$, PF$00072$ and PF$00096$) or  remain unchanged (for PF$00013$, PF$00014$, PF$00076$, PF$00887$ and PF$13354$) after sampling at $T^\mathrm{best}$.
       The only exceptions are the protein families PF$00041$ and (especially) PF$00153$, where the sampling worsens the test error obtained after the training with Adam and early stopping.

\begin{figure}
            \centering
            \includegraphics[width=\linewidth]{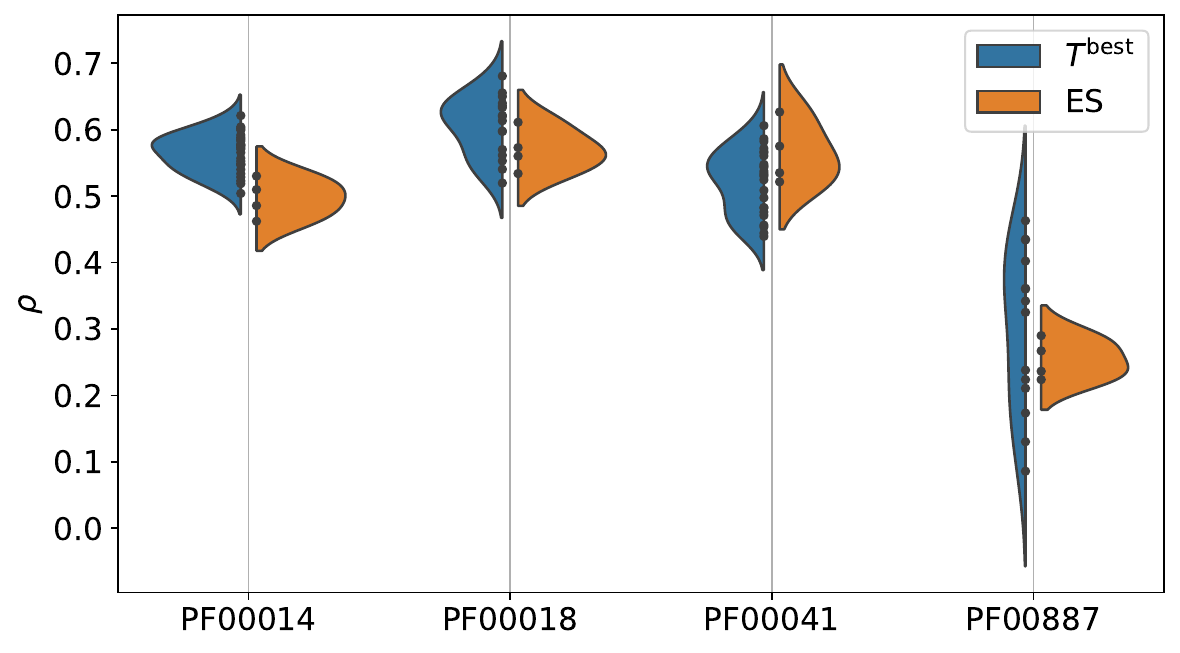}
            \caption{
            The distribution of the Spearman correlation coefficient computed between experimental and predicted $\Delta \Delta G$, for models sampled at $T^{\mathrm{best}}$ (blue curves) and for early-stopped solutions (orange curves), for four sequences belonging to different protein families (from left to right, PF$00014$, PF$00018$, PF$00041$ and PF$00887$).
            The correlation coefficient associated to each model (either sampled or trained) is shown as a dot in the respective distribution.
            }
            \label{fig:Pfam_ddGs_corr}
        \end{figure}

       To understand if the dominant feature learned by the network is still thermodynamic stability, we investigated the ability of the trained and sampled models to predict the free-energy variation $\Delta \Delta G$ caused by single-site mutations within a specific sequence, which are known for some proteins as the result of protein--engineering experiments. For four such proteins (belonging to the families PF$00014$ \cite{Yu1995ContributionMutagenesis}, PF$00018$ \cite{Grantcharova1998}, PF$00041$ \cite{Cota2000TwoStability} and PF$00887$ \cite{Kragelund1999}, respectively), we predicted the free energy change $\Delta\Delta G^{\mathrm{pred}}$ from the variation of the logarithmic output score (Eq. \ref{app_eq:ddG_pred} in the Appendix). We then calculated the Spearman correlation coefficient between the experimental values ($\Delta\Delta G^{\mathrm{exp}}$) and the predicted values ($\Delta\Delta G^{\mathrm{pred}}$), obtained by either ES solutions or models sampled at $T^\mathrm{best}$ (Fig. \ref{app_fig:Pfam_ddGs_scatterplot}). In three cases (Fig. \ref{fig:Pfam_ddGs_corr}), the correlation coefficient obtained from the sampled models is larger than 0.5, suggesting that the model takes the energetic properties of the protein into account. In the fourth case (PF$00887$), the results of the sampled models vary considerably, with some unable to capture the interactions between amino acids.
       
       Finally, in three out of four cases, the distribution of the correlation coefficients for the sampled models is peaked at larger values than those obtained by gradient minimization.
       The only exception is the sequence belonging to the PF$00041$ protein family, which is also one of the two datasets over which the sampling slightly worsens the model performance in terms of test error (see Fig. \ref{fig:Pfam_thermodynamics}).
        
\section{Discussion and Conclusions}

    Transformers are a widespread tool used for many tasks, most notably in large language models and protein structure prediction. They are extremely powerful and have surpassed multi--layer perceptrons and recurrent networks in many areas, yet a deep understanding of how they work is lacking. In this regard, statistical mechanics has been useful in elucidating the working mechanisms of several artificial neural network architectures \cite{Seung1992StatisticalExamples,Monasson1994DomainsNetworks,Bahri2020,Baldassi2015,Baldassi2016UnreasonableSchemes,Rocks2022MemorizingModels,Becker2020GeometryNetworks,Zambon2025SamplingNetwork}.
    
    In order to clarify how transformers function in protein structural prediction, we explored the parameter space of transformers that were trained to recognize protein sequences. Our goal is to understand why transformers are so efficient by exploring the low-loss manifold of the system. One way to do this is to sample the parameter space at a low temperature in the canonical ensemble. This transforms the problem into the realm of statistical mechanics, allowing us to use computational algorithms developed in this field over the last fifty years. Specifically, we employed a type of molecular dynamics in the parameter space driven by approximate Langevin equations, which are highly effective for sampling artificial neural networks \cite{Zambon2026ControlledMinibatches}.
    
    We studied small transformers made of either one or four blocks for the simple task of learning the statistical properties of amino acid sequences. We used for training and test either synthetic sequences, generated by energy minimization on a specific native conformation, having the only property of folding stably \cite{Shakhnovich1993,Franco2019}, and real sequences, belonging to ten different protein families. This allowed us to explore the space of parameters in detail under different conditions without being strongly limited by computational constraints. Of course, we cannot yet determine how our results scale for larger transformers and more ambitious tasks, such as learning the structure of all known proteins.

    An interesting feature that emerges from our calculations is a wide range of intermediate temperatures at which the generalization capability of the transformer is particularly high, as testified by lower values in test error. Basically the generalization error is a non--monotonic function of the temperature both for synthetic and real proteins. At lower temperatures, the loss and the training error are smaller, but also the ability to generalize is smaller. In this respect, transformers behave quite differently from pooled feedforward networks with similar number of parameters and similar architecture, but lacking the attention mechanism. Pooled feedforward networks display a sharp transition in the average loss (and consequently in the training error) with respect to temperature from a low--temperature state with good training but bad generalization accuracy (i.e., the network interpolates the data), to an higher--temperature state where the training error increases drastically. This behavior seems to affect pooled feedforward networks independently of their detailed architecture and task \cite{Zambon2025SamplingNetwork}. On the contrary, in transformers the transition is smooth, and does not appear to sharpen increasing the size of the network, generating a wide range of temperatures where the training error is still small enough and where generalization is more efficient.
    
    This behavior explains why the typical large language models work so well despite being largely under--trained \cite{Hoffmann2022TrainingModels}. Due to the huge number of parameters, it would anyway be impossible to minimize their loss to a state corresponding to the low-temperature states observed in our system. Instead, their optimization approaches the wide intermediate-temperature region that is associated with better generalization properties.

    From the sampling at intermediate temperatures we also learn that most layers of the transformer can be highly variable from model to model. Similarly, the pooled feedforward network that we use as a control displays similar patters, therefore this behavior does not depend on the attention mechanism. 
    A group of particularly conserved parameters are the ones associated with normalization layers, whose value distribution assumes a bimodal shape with one peak centered around zero during samplings in the intermediate temperature range. Since these vectors have dimension $d$ equal to the embedding size, the peak centered around higher values identifies the minimal dimension needed to obtain good performances on the masked token prediction task. 

    In the minimal model obtained from decreasing the size of the embedding, the parameters of several layers become significantly more conserved along the sampling compared to those of the larger model. The matrices defining the embedding, the query/key, the value, the normalization and the output have little freedom to vary in the one--block transformer, and somewhat more in the multi--block architecture. Only the biases are really free to vary.

    The fact that the attention matrix is similar to the contact map of proteins is a useful byproduct of transformers trained with amino acid sequences \cite{Rao2020TransformerLearners}. The sampling at intermediate temperatures produce attention matrices that are more similar to the experimental contact map than loss optimization. This is true for both transformer architectures we studied, although in the case of multi--blocks one faces the problem of identifying {\it a priori} what block has the most informative attention matrix.
    Interestingly, transformers sampled at temperatures higher than those optimal for training and testing sequences, produce attention matrices which are more similar to the experimental contact map. Over--trained solutions seem to perform better in the prediction of native contacts than early--stopped ones. The reduction of the dimension of the embedding that have little effect on the ability of the transformer to learn protein sequences, decreases the ability to predict the contact map. The misalignment between conditions that favors sequence learning and those that favors the prediction of the contact map is not unexpected. As a matter of fact, the network is not trained to reproduce the latter. We can speculate that at higher temperatures or for larger embeddings, it is more difficult for the transformer to predict protein sequences out of noise, and this requires a more selective representation of the correlations between amino acids; the attention matrix is then expected to be cleaner and more similar to the contact map.

    The results obtained with real protein families are as a rule in agreement with those obtained with synthetic sequences. For the formers we had also the possibility to compare the energetic effect of mutations with the experimental values, which are an independent set of data available for some of them. The fair match we obtained suggests not only that the sampled models are learning correctly the properties of the proteins, but also that they are dominated by thermodynamic requirements.
    
    Finally, we suggest that the Langevin strategy we used in this study can be an efficient tool for optimization of large language models, as it provides configurations with better predictive capabilities for both problems, i.e., masked token prediction and native contacts prediction. A proper tuning of the temperature in the intermediate-to-high range could help find models with better generalization properties.

    \vspace{0.5cm}
    \centerline{\bf\small DATA AVAILABILITY}
    \vspace{0.2cm}

    The codes \cite{linkCode} used for the sampling can be downloaded at \url{https://github.com/guidotiana/PseudoLangevin}. The data \cite{linkData} generated by our sampling algorithm are available at \url{https://doi.org/10.13130/RD_UNIMI/J1TOFK}.

\acknowledgements{We would like to thank Riccardo Zecchina for his valuable discussions and suggestions.}

\bibliography{references}

\clearpage

\appendix
\renewcommand\thefigure{\thesection\arabic{figure}}    
\renewcommand\thetable{\thesection\arabic{table}}    

\setcounter{figure}{0}    
\setcounter{table}{0}

\section{The Pfam datasets}
\label{app_sect:Pfam_dataset}

    The ten datasets of real protein sequences were obtained from the complete, annotated Pfam sequence alignment using the following procedure. 
    
    First, we identified all sequences belonging to the same selected protein family. From these sequences, we selected a representative one. We then removed all sites containing a gap in the reference sequence from each sequence in the alignment. Then, we discarded all those with more than $10\%$ of their length composed of gaps after the removal of the sites. Finally, we clustered the remaining sequences using MMseqs2 \cite{Steinegger2017MMseqs2Sets} with a similarity threshold of $70\%$, and then selected a single representative sequence from each cluster.

    In this way, we created protein datasets of fixed length (as in the case of the synthetic dataset), discarding sequences with a high proportion of gap sites and also sequences that were too similar to each other. The selected protein families and the code of the associated representative sequence can be found in Tab. \ref{app_tab:train_insight_Pfam}.

\section{The models}
\label{app_sect:models}

    The architectures considered in this study follow a common three-stage structure: an input construction stage, an encoder, and a final output projection layer. The input and output components are identical for the two network families studied here, while the internal encoder differs between transformer-based and pooled feedforward models.

    \subsection{Embedding}
    \label{app_sect:models_embedding}
        All models operate on sequences $\boldsymbol{x}=(x_1,\dots,x_n)$ of $n$ discrete tokens drawn from a finite vocabulary. The latter is composed of $20$ amino acid tokens and a special masking symbol added to represent missing or hidden tokens in the input sequence. Its cardinality is therefore $|V|=21$.
        
        In the input stage, each token $x_i \in \{1,\dots,|V|\}$ is represented by a one-hot vector $\boldsymbol{e}_{x_i} \in \{ 0,1\}^{|V|}$ whose $x_i$-th component is equal to 1 and all other components are 0, and then is mapped to a continuous vector representation through an embedding layer. Given an embedding matrix $E\in\mathbb{R}^{|V|\times d}$, the embedded representation of the token is obtained as
        \begin{equation}
        \boldsymbol{x}^{(E)}_i = E^\top \boldsymbol{e}_{x_i} \in \mathbb{R}^d,
        \end{equation}
        producing a matrix sequence representation as $X^{(E)} \in\mathbb{R}^{n\times d}$, where $n$ denotes the sequence length and $d$ the embedding size.
        After the embedding, positional information is added through standard sinusoidal positional encoding \cite{Vaswani2023AttentionNeed},
        \begin{equation}
        P^{(z)}_{k,2r} = \sin\!\left(\frac{k}{10000^{2r/z}}\right), \quad
        P^{(z)}_{k,2r+1} = \cos\!\left(\frac{k}{10000^{2r/z}}\right),
        \end{equation}
        with $P^{(z)} \in \mathbb{R}^{n \times z}$. This yields the common encoder input
        \begin{equation}
        \tilde{X} = X^{(E)} + P^{(d)}.
        \label{app_eq:enc_input}
        \end{equation}

    \subsection{Compact transformer encoder}
    \label{app_sect:models_transfenc}
        The first encoder considered in this work is a simplified compact transformer inspired by the standard architecture in ref. \cite{Thickstun_web}. 
        
        Let $\tilde{X} \in \mathbb{R}^{n \times d}$ denote the input representation of a given encoder block. The self-attention mechanism computes interactions between embedded tokens through a bilinear form
        \begin{equation}
        A = \mathrm{softmax}\!\left(\frac{\tilde{X} W^{(QK)} \tilde{X}^\top}{\sqrt{d}}\right)
        \in \mathbb{R}^{n \times n},
        \label{app_eq:attention}
        \end{equation}
        where the softmax is applied row-wise, and where the matrix 
        $W^{(QK)} \equiv W^{(Q)}(W^{(K)})^T \in \mathbb{R}^{d \times d}$ represents the combined query–key transformation.
        
        The attention weights are then used to aggregate token information as follows
        \begin{equation}
        U^{(0)} = A \tilde{X} W^{(VC)} \in \mathbb{R}^{n \times d}
        \label{app_eq:Wvc}
        \end{equation}
        with $W^{(VC)} \equiv W^{(V)}W^{(C)} \in \mathbb{R}^{d \times d}$ as the combined value-output projection matrix.
        
        This contribution is added to the residual branch and normalized
        \begin{equation}
        U = \mathrm{Norm^{(1)}}(\tilde{X} + U^{(0)}),
        \end{equation}
        where ``$\mathrm{Norm}$" denotes layer normalization applied independently to each token representation. 
        Given a generic matrix $V \in \mathbb{R}^{a \times b}$, the normalization is performed row-wise, i.e. independently for each vector $V_i \in \mathbb{R}^b$. 
        For the $i$-th row, we compute
        \begin{equation}
        \mu_i = \frac{1}{b}\sum_{j=1}^{b} V_{ij},
        \qquad
        \sigma_i^2 = \frac{1}{b}\sum_{j=1}^{b} (V_{ij}-\mu_i)^2 ,
        \end{equation}
        and the normalization layer output is then
        \begin{equation}
        \mathrm{Norm}(V)_{ij} = \frac{V_{ij}-\mu_i}{\sigma_i} \left(\gamma_{\mathrm{weight}}\right)_{j} + \left(\gamma_{\mathrm{bias}}\right)_{j} ,
        \end{equation}
        with $\boldsymbol{\gamma}_{\mathrm{weight}},\boldsymbol{\gamma}_{\mathrm{bias}} \in \mathbb{R}^b$ being learnable vectors applied element-wise along the feature dimension.
        
        The subsequent feedforward stage expands the representation to a higher-dimensional space
        \begin{equation}
        H = U W^{(1)} + \boldsymbol{b}^{(1)} \, \in \mathbb{R}^{n \times m}
        \end{equation}
        where $m$ represents the size of the hidden feedforward layer, and where $W^{(1)} \in \mathbb{R}^{d \times m}$ and $\boldsymbol{b}^{(1)} \in \mathbb{R}^{m}$ are the associated weight matrix and bias vector respectively.
        
        After applying a non-linearity, an additional positional contribution is introduced in the hidden representation
        \begin{equation}
        \tilde{H} = \mathrm{ReLU}(H) + P^{(m)}.
        \label{app_eq:extra_position_enc_transformer}
        \end{equation}
        where the function $\mathrm{ReLU}(x) = \max(0, x)$ is applied element-wise.
        
        The representation is then projected back to the model dimension
        \begin{equation}
        Z^{(0)} = \tilde{H} W^{(2)} + \boldsymbol{b}^{(2)} \, \in \mathbb{R}^{n \times d}
        \end{equation}
        where $W^{(2)} \in \mathbb{R}^{m \times d}$ and $\boldsymbol{b}^{(2)} \in \mathbb{R}^{d}$ are the weight matrix and the bias vector of the second hidden feedforward layer of size $d$.
        
        Lastly, the output of the encoder block is obtained through a second residual connection followed by a normalization
        \begin{equation}
        Z = \mathrm{Norm^{(2)}}(U + Z^{(0)}) \, \in \mathbb{R}^{n \times d}
        \end{equation}
        which represents the contextualized token representation produced by the block.
        
        The encoder described above corresponds to a single transformer block. To increase the expressive capacity of the model, multiple identical blocks can be composed sequentially. Let $L$ denote the depth of the encoder, or the number of stacked blocks. Given an input representation, the output of each block is used as the input for the next block. Thus, the full encoder is obtained by composing L transformations. The stacked blocks share the same functional form, yet they have separate, trainable parameters.

    \subsection{Feed-forward encoder}
    \label{app_sect:models_ffenc}
        The second type of encoder can be interpreted as a feed-forward network with an organization similar to that of a transformer.
        It has the same residual structure as the transformer block but replaces the attention mechanism with a simpler global context aggregation step. Rather than computing pairwise token interactions through an attention matrix, contextual information is introduced via a pooling operation over the sequence. This difference makes the model a useful baseline for isolating the contribution of self-attention and, more generally, of explicit contextualization mechanisms.
        
        Starting from the shared input representation $\tilde{X}$ (see Eq. \eqref{app_eq:enc_input}), the first step consists of applying layer normalization independently to each token representation,
        \begin{equation}
        U = \mathrm{Norm^{(1)}}(\tilde{X}).
        \end{equation}
        This normalization stabilizes the representation and ensures that feature magnitudes are comparable across positions. 
        The next step is to construct a simple form of contextual information shared by the whole sequence, extracting a global summary from the visible tokens only.
        
        More precisely, given a binary visibility mask $\chi\in\{0,1\}^n$, where $\chi_i=0$ for masked positions and $1$ otherwise, we define the global context vector $g$ as the column-wise average of the representations of the visible tokens,
        \begin{equation}
        g = \frac{1}{\sum_{i=1}^{n}\chi_i}\sum_{i=1}^{n}\chi_i\,U_i
        \in \mathbb{R}^{d}.
        \end{equation}
        The vector $g$ provides a coarse summary of the sequence based only on unmasked tokens, and it is then broadcast to all positions,
        \begin{equation}
        G = \mathbf{1}_n g^\top
        \in \mathbb{R}^{n\times d}.
        \end{equation}
        
        The local representation $U$ and the global broadcast context $G$ are finally concatenated along the feature dimension,
        \begin{equation}
        S = [\,U \;\; G\,] 
        \in \mathbb{R}^{n\times 2d}.
        \end{equation}
        Through this construction, each token receives both its local features and a shared global summary of the sequence. This pooling mechanism provides a simple form of contextualization that replaces the pairwise interactions produced by the attention matrix in the transformer encoder (see Eq. \eqref{app_eq:attention}).
        
        Then, similarly to the transformer encoder, a two-layer feedforward transformation is applied. The first linear projection expands the representation to an internal dimension $m$,
        \begin{equation}
        H = S W^{(1)} + \boldsymbol{b}^{(1)} \, \in \mathbb{R}^{n\times m}
        \end{equation}
        where $W^{(1)}\in\mathbb{R}^{2d\times m}$ and $\boldsymbol{b}^{(1)}\in\mathbb{R}^{m}$ are the weight matrix and bias vector of the first hidden layer.
        
       A non-linearity is applied and then a positional contribution is introduced in the hidden representation,
        \begin{equation}
        \tilde{H} = \mathrm{ReLU}(H) + P^{(m)}.
        \label{app_eq:extra_position_enc_feedforward}
        \end{equation}
        
        The hidden representation is then processed by a second linear transformation within the internal space,
        \begin{equation}
        F = \mathrm{ReLU}(\tilde{H} W^{(2)} + \boldsymbol{b}^{(2)}) \, \in \mathbb{R}^{n\times m}
        \end{equation}
        where once more $W^{(2)}\in\mathbb{R}^{m\times m}$ and $\boldsymbol{b}^{(2)}\in\mathbb{R}^{m}$ are the weight matrix and the bias vector of the second feedforward layer.
        
        In parallel, a residual projection of the normalized input is computed,
        \begin{equation}
        R = U W^{(r)} + \boldsymbol{b}^{(r)} \, \in \mathbb{R}^{n\times m}
        \end{equation}
        with $W^{(r)}\in\mathbb{R}^{d\times m}$ and $\boldsymbol{b}^{(r)}\in\mathbb{R}^{m}$. 
        The encoder output is then obtained by combining the two streams and applying a normalization layer
        \begin{equation}
        Z = \mathrm{Norm^{(2)}}(R + F).
        \end{equation}
        
        In contrast to the transformer architecture, this feed-forward encoder is implemented using a single block and no stacking is considered. Therefore no depth parameter $L$ is introduced for this model.

    \subsection{Output}
    \label{app_sect:models_output}
        After the input construction stage, the encoder transforms the input representation into contextual features that capture dependencies between sequence positions. The encoder output is a sequence of hidden representations
        \begin{equation}
        Z \in \mathbb{R}^{n \times h},
        \end{equation}
        where $h$ denotes the dimensionality of the internal representation. The precise transformation that produces $Z$ depends on the specific encoder architecture described in the previous sections.
        
        In the final stage, the encoded representation is projected back into the vocabulary space through a linear transformation,
        \begin{equation}
        \hat{\boldsymbol{y}}_i = (W^{(O)})^\top Z_i + \boldsymbol{b}^{(O)},
        \end{equation}
        where $W^{(O)} \in \mathbb{R}^{h \times |V|}$ and $\boldsymbol{b}^{(O)} \in \mathbb{R}^{|V|}$. The resulting vector $\hat{\boldsymbol{y}}_i \in \mathbb{R}^{|V|}$ contains the model scores associated with each vocabulary token for the site $i$ along the input sequence.

    \subsection{Details and hyperparameters of our models}
    \label{app_sect:models_summary}
        All models share the same input embedding and output projection layers introduced in the previous section and differ only in the internal encoder. This makes it possible to compare architectures while keeping the input representation and prediction head fixed.
        
        We consider two transformer-based models, obtained by stacking identical encoder blocks with depth $L \in \{ 1, 4\}$.
        In both cases, the embedding dimension is fixed to $d=64$ and the internal feed-forward width to $m=4d=256$. These two configurations allow us to compare a shallow and a deeper transformer while preserving the same block structure.

        In addition, we consider two smaller transformer-based model, again with one module (with $d=26$ and $m=4d=104$) and four modules (with $d=14$ and $m=4d=56$), whose embedding size $d$ is derived from the Boltzmann sampling of their respective larger counterparts (see Sect. \ref{sect:var_par}).
        
        As a baseline, we also consider a feed-forward-based model.
        For this model, the dimensions are fixed to $d = 64$ and $m=320$. The size of the first feedforward layer $m$ is chosen so that the total number of trainable parameters is comparable to that of the bigger transformer (i.e. $L=4$ and $d=64$). In this way, the comparison isolates the effect of attention while keeping the overall parameter scale similar.
        The resulting model configurations are summarized in Table~\ref{app_tab:model_configs}.

        \begin{table}[h]
        \centering
        \begin{tabular}{lcccc}
        \hline
        \makecell{Encoder \\ Type} & $L$ & $d$ & $m$ & \makecell{Params \\ $[\times 10^3]$} \\ \hline
        Transformer & 1 & 26 & 104 & 8    \\ 
        Transformer & 1 & 64 & 256 & 44   \\
        Transformer & 4 & 14 & 56  & 9    \\
        Transformer & 4 & 64 & 256 & 169  \\ 
        Feed-forward & - & 64 & 320 & 174 \\ \hline
        \end{tabular}
        \caption{Model configurations and number of trainable parameters.}
        \label{app_tab:model_configs}
        \end{table}

        Lastly, it should be noted that both encoder architectures have slight modifications to their typical compositions. For example, in the transformer-based encoder, we merged the query and key matrices (see Eq. \eqref{app_eq:attention}) and the value and output matrices (see Eq. \eqref{app_eq:Wvc}). Additionally, we applied an extra positional encoding to the output of the first hidden layer for both architectures (see Eq. \eqref{app_eq:extra_position_enc_feedforward} and Eq. \eqref{app_eq:extra_position_enc_transformer}).
        These modifications were made to eliminate all possible permutation symmetries within the studied models, without strongly affecting their learning capabilities (see Tab. \ref{app_tab:train_insight}).

\section{Observables}
\label{app_sect:observables}
    Let's consider a neural network model among the ones previously described (Appendix \ref{app_sect:models_summary}) and let's call $\boldsymbol{w}$ the weight vector representing all its parameters. The quality of the prediction of the NN over a dataset $\mathcal{D}$ (as the ones defined in Sect. \ref{sect:synthetic_dataset} and Sect. \ref{sect:Pfam_dataset}) is evaluated using two quantities: the cross--entropy loss function and the error function.
        
    For a given pair $\left( \boldsymbol{x}, \boldsymbol{y} \right)$, let \(\mathcal{M}\subset\{1,\dots,n\}\) denote the set of associated masked indices, and let \(\hat{\boldsymbol{y}}_i\in\mathbb{R}^{|V|}\) be the vector of output scores for the masked position \(i\) (see Appendix \ref{app_sect:models_output}). 
        
    For the selected input-label pair, the cross--entropy loss is defined as
    \begin{equation}
        \ell_{\mathrm{CE}} \left( \boldsymbol{w} | (\boldsymbol{x}, \boldsymbol{y}) \right)
        =
        -
        \frac{1}{|\mathcal{M}|}
        \sum_{i\in\mathcal{M}}
        \sum_{v=1}^{|V|}
        e_{y_i, v} ,
        \log\big(\mathrm{softmax}(\hat{\boldsymbol{y}}_i)_v\big),
    \end{equation}
    where $\boldsymbol{e}_{y_i} \in \{0, 1\}^{|V|}$ denotes the one--hot encoded vector representing the true amino acid at masked site \(i\). Analogously, the global loss function is obtained by averaging over all the pairs, that is
    \begin{equation}
    \mathcal{L}(\boldsymbol{w} | \mathcal D)
    =
    \frac{1}{P}
    \sum_{\mu=1}^{P}
    \ell_{\mathrm{CE}} \left( \boldsymbol{w} | (\boldsymbol{x}^\mu, \boldsymbol{y}^{\mu}) \right),
    \label{app_eq:loss}
    \end{equation}
    which in our case coincides with the potential energy $U(\boldsymbol{w} | \mathcal D)$ associated with the weight vector $\w$, as no regularization terms were added.

    From the same $\hat{\boldsymbol{y}}_i$ vector, we can define the most probable amino acid at the masked position $i$ as
    \begin{equation}
    \hat{c}_i = \mathrm{argmax}_v \left[ \mathrm{softmax}(\hat{\boldsymbol{y}}_i)_v \right].
    \end{equation}
    with $\hat{c}_i \in V$ being a token from the vocabulary. The masked--token accuracy evaluated on the $\left( \boldsymbol{x}, \boldsymbol{y} \right)$ pair is then
    \begin{equation}
    a \left( \boldsymbol{w} | (\boldsymbol{x}, \boldsymbol{y}) \right)
    =
    \frac{1}{|\mathcal{M}|}
    \sum_{i\in\mathcal{M}}
    \mathbf{1}\!\left[\hat{c}_i = y_i\right].
    \end{equation}
    As before, the global error function can be computed by averaging over all the input-label couples in the dataset,
    \begin{equation}
    \epsilon \left( \boldsymbol{w} | \mathcal{D} \right)
    =
    \frac{1}{P}
    \sum_{\mu=1}^{P}
    \left[ 1 - a\left( \boldsymbol{w} | (\boldsymbol{x}^\mu, \boldsymbol{y}^\mu) \right) \right].
    \label{app_eq:error}
    \end{equation}

    Lastly, in Sect. \ref{sect:Pfam}, we characterized the capability of a given model to predict the free energy variation in a sequence after a single-site mutation, that is, $\Delta \Delta G^{\mathrm{pred}}$. Let's consider a sequence $\boldsymbol{x}$ masked on site $j$, and a single-site mutation on the same site, that is, $x_j=\sigma \rightarrow \xi$. Then, the predicted free energy variation deriving from such mutation is defined as follows,
    \begin{equation}
    \Delta \Delta G^{\mathrm{pred}} \left( \boldsymbol{w} | \boldsymbol{x}, \sigma, \xi, j \right)
    =
    - \log\big( \frac{\mathrm{softmax}(\hat{\boldsymbol{y}}_j)_{\xi}}{\mathrm{softmax}(\hat{\boldsymbol{y}}_j)_{\sigma}} \big)
    \label{app_eq:ddG_pred}
    \end{equation}
    where, as before, $\hat{\boldsymbol{y}}_j$ is the vector of output scores for the site $j$ given the masked sequence $\boldsymbol{x}$ as input.

\section{Training algorithms}
\label{app_sect:training}

    All models are trained using standard optimization techniques commonly adopted for transformer-based architectures ~\cite{Vaswani2023AttentionNeed}. 
    Both the transformer-based and the feedforward models are optimized with Adam, using momentum parameters \((\beta_1,\beta_2)=(0.9,0.98)\) and \(\epsilon=10^{-9}\). 
    The learning rate follows the standard schedule, consisting of a linear warm-up phase followed by inverse square-root decay. At optimization step \(t\), the learning rate is
    \begin{equation}
    \eta(t) = d^{-1/2}\;
    \min\!\left(t^{-1/2},\; ts^{-3/2}\right),
    \end{equation}
    where \(d\) denotes the embedding dimension and \(s=4000\) is the number of warm-up steps.
        
    A dropout rate of \(p_{\mathrm{drop}}=0.1\) is used throughout, following the placement adopted in the original transformer: dropout is applied to the input embeddings, to the outputs of attention and feed-forward sublayers. No additional explicit regularization term is introduced during training.
        
\begin{figure}
        \centering
        \includegraphics[width=\linewidth]{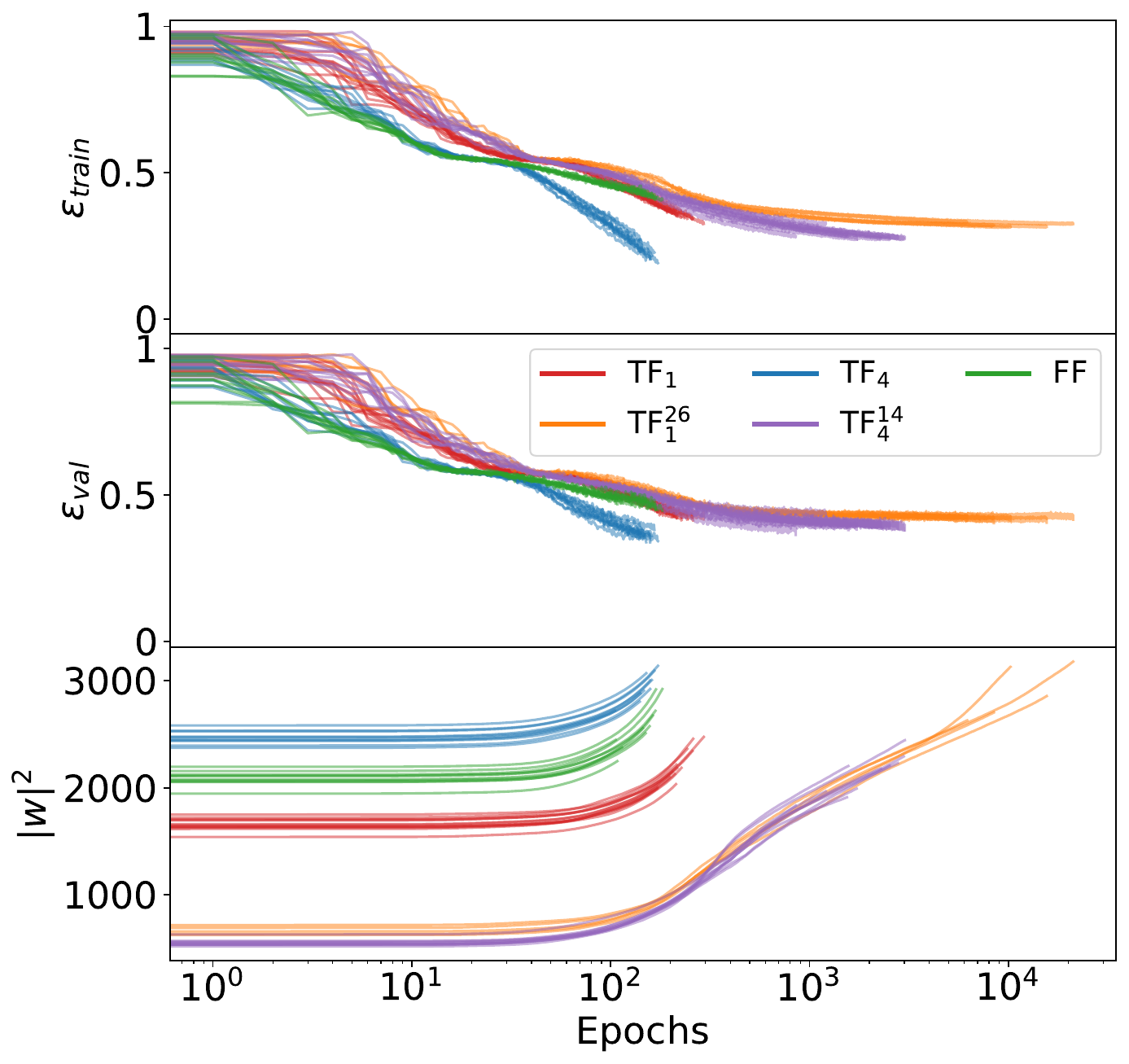}
        \caption{
        The training error $\epsilon_{\mathrm{train}}$ (upper panel), the validation error $\epsilon_{\mathrm{val}}$ (middle panel), and the squared norm $|\w|^2$ (lower panel) as functions of the training epoch during several first--stage trainings for the $\mathrm{TF}_{1}$ model (red curves), the $\mathrm{TF}_{1}^{26}$ model (orange curves), the $\mathrm{TF}_{4}$ model (blue curves), the $\mathrm{TF}_{4}^{14}$ model (purple curves) and the $\mathrm{FF}$ model (green curves).
        }
        \label{app_fig:first_stage}
    \end{figure}

    We adopt a two-stage training protocol. 
    In the first stage, multiple independent instances of each model are trained with early--stopping based on validation cross-entropy. 
    From the ensemble of trained networks, we compute the empirical distribution of the squared norm of the parameter vectors, \(|\boldsymbol{w}|^2\), and estimate its mean value, with $\boldsymbol{w}$ denoting the full vector of trainable parameters of the network. For each architecture, the first stage is repeated over ten independent runs with different random initializations.
    The resulting training dynamics and norm evolution across independent runs for the synthetic dataset are shown in Fig.~\ref{app_fig:first_stage}. 
        
    In the second stage, we train three final reference models for each architecture type while explicitly monitoring \(| \boldsymbol{w}|^2\). 
    Training is stopped as soon as the squared norm reaches the mean value estimated in the first stage for the corresponding model class, to save the reached configuration (ES models). The ES solutions are then employed as starting points in the samplings at fixed norm performed using the Constrained pseudo-Langevin algorithm (see Appendix \ref{app_sect:sampling}).
    
    Additionally, for the synthetic dataset, each training procedure continues without any norm constraints up until both the cross--entropy function and the error function evaluated on the train set saturate (OT models). 
    
    The performances of trained models for each architecture are summed up in Tab. \ref{app_tab:train_insight} (for the synthetic sequence datasets) and in Tab. \ref{app_tab:train_insight_Pfam} (separately for each protein family studied).
    
    \begin{table*}
        \centering
        \renewcommand{\arraystretch}{1.2}
        \setlength{\tabcolsep}{4pt}
        
        \begin{tabular}{
        lcccc|cc
        S[table-format=1.2] @{\,\(\pm\)\,} S[table-format=1.3]
        S[table-format=1.2] @{\,\(\pm\)\,} S[table-format=1.3]
        }
        \hline
        \makecell{Encoder \\ Type} & $L$ & $d$ & $m$ & \makecell{Params \\ $[\times 10^3]$} & ES & Epochs
        & \multicolumn{2}{c}{$\langle \epsilon_{\mathrm{train}}\rangle$}
        & \multicolumn{2}{c}{$\langle \epsilon_{\mathrm{test}}\rangle$} \\
        \hline
        \makecell{Transformer}  & 1  & 26 & 104 & 8   & $\times$     & 100k & 0.31 & 0.006 & 0.40 & 0.006 \\
        \makecell{Transformer}  & 1  & 26 & 104 & 8   & $\checkmark$ & 10k  & 0.32 & 0.006 & 0.39 & 0.005 \\
        \makecell{Transformer}  & 1  & 64 & 256 & 44  & $\times$     & 500k & 0.02 & 0.003 & 0.43 & 0.004 \\
        \makecell{Transformer}  & 1  & 64 & 256 & 44  & $\checkmark$ & 240  & 0.35 & 0.02  & 0.40 & 0.008 \\
        \makecell{Transformer}  & 4  & 14 & 56 & 9 & $\times$     & 100k  & 0.22  & 0.014  & 0.36 & 0.16  \\
        \makecell{Transformer}  & 4  & 14 & 56 & 9 & $\checkmark$     & 2k  & 0.40  & 0.012  & 0.37 & 0.006  \\
        \makecell{Transformer}  & 4  & 64 & 256 & 169 & $\times$     & 1k   & 0.00 & 0.00  & 0.36 & 0.005  \\
        \makecell{Transformer}  & 4  & 64 & 256 & 169 & $\checkmark$ & 170  & 0.22 & 0.04  & 0.35 & 0.016  \\
        \makecell{Feedforward}  & -- & 64 & 320 & 174 & $\times$     & 2k   & 0.00 & 0.00  & 0.50 & 0.012 \\
        \makecell{Feedforward}  & -- & 64 & 320 & 174 & $\checkmark$ & 170  & 0.42 & 0.012 & 0.46 & 0.006 \\
        \hline
        \end{tabular}
        \caption{
        Results of the second-stage trainings for ES ($\checkmark$) and OT ($\times$) solutions, with the associated number of training epochs (Epochs) and their average performance on the train and test synthetic datasets evaluated through the error function ($\langle \epsilon_{\mathrm{train}} \rangle$ and $\langle \epsilon_{\mathrm{test}} \rangle$ respectively).
        }
        \label{app_tab:train_insight}
    \end{table*}

    \begin{table*}
        \centering
        \renewcommand{\arraystretch}{1.2}
        \setlength{\tabcolsep}{4pt}
        
        \begin{tabular}{
        lcccc|ccc
        S[table-format=1.3] @{\,\(\pm\)\,} S[table-format=1.4]
        S[table-format=1.3] @{\,\(\pm\)\,} S[table-format=1.4]
        }
        \hline
        \makecell{Encoder \\ Type} & $L$ & $d$ & $m$ & \makecell{Params \\ $[\times 10^3]$} & Pfam & PDB & Epochs
        & \multicolumn{2}{c}{$\langle \epsilon_{\mathrm{train}}\rangle$}
        & \multicolumn{2}{c}{$\langle \epsilon_{\mathrm{test}}\rangle$} \\
        \hline
        \makecell{Transformer}  & 4  & 64 & 256 & 169 & PF$00013$ & 4QVZ & 40   & 0.586 & 0.004  & 0.619 & 0.005  \\
        \makecell{Transformer}  & 4  & 64 & 256 & 169 & PF$00014$ & 1PIT & 50   & 0.491 & 0.003  & 0.515 & 0.007 \\
        \makecell{Transformer}  & 4  & 64 & 256 & 169 & PF$00018$ & 1SHG & 60   & 0.502 & 0.007  & 0.560 & 0.008  \\
        \makecell{Transformer}  & 4  & 64 & 256 & 169 & PF$00041$ & 1FNF & 70   & 0.565 & 0.003  & 0.602 & 0.0014  \\
        \makecell{Transformer}  & 4  & 64 & 256 & 169 & PF$00072$ & 3CHY & 100  & 0.544 & 0.0005 & 0.549 & 0.0007 \\
        \makecell{Transformer}  & 4  & 64 & 256 & 169 & PF$00076$ & 1HA1 & 60   & 0.573 & 0.004  & 0.616 & 0.002  \\
        \makecell{Transformer}  & 4  & 64 & 256 & 169 & PF$00096$ & 1SP1 & 30   & 0.545 & 0.0008 & 0.546 & 0.0009  \\
        \makecell{Transformer}  & 4  & 64 & 256 & 169 & PF$00153$ & 1OKC & 130  & 0.503 & 0.005  & 0.568 & 0.0009 \\
        \makecell{Transformer}  & 4  & 64 & 256 & 169 & PF$00887$ & 2ABD & 80   & 0.484 & 0.028  & 0.539 & 0.005  \\
        \makecell{Transformer}  & 4  & 64 & 256 & 169 & PF$13354$ & 1HTZ & 120  & 0.447 & 0.005  & 0.538 & 0.0011 \\
        \hline
        \end{tabular}
        \caption{
        Results of the second-stage trainings for $\mathrm{TF}_4$ ES solutions on each Pfam dataset (Pfam), with the associated number of training epochs (Epochs) and their average performance on the train and test sets evaluated through the error function ($\langle \epsilon_{\mathrm{train}} \rangle$ and $\langle \epsilon_{\mathrm{test}} \rangle$ respectively). Each protein family code is associated with the PDB code of the reference sequence used to generate the dataset.
        }
        \label{app_tab:train_insight_Pfam}
    \end{table*}

\section{The sampling algorithm}
\label{app_sect:sampling}


        The pseudo--Langevin (pL) algorithm \cite{Zambon2026ControlledMinibatches} is an approximate integrator of the Langevin stochastic differential equations of motion. It produces a dynamics in the phase space $\left( \w, \p \right)$, with $\p$ being the momenta associated with the weight vector $\w$, which efficiently samples the Boltzmann equilibrium distribution at a certain temperature $T$.

        We modified the pL sampling scheme, in order to perform a norm--constrained dynamics in the space of the weights of the neural network. Imposing a constant norm for the vector of parameters $\w$ at all times during integration can be translated into two equations, that is one for the holonomic constraint 
        \begin{equation}
            C(\w) = |\w|^2 - Q= 0,
            \label{app_eq:constraint_w}
        \end{equation}
        and one for its time derivative
        \begin{equation}
            \dot{C}(\w) = 2 \w \dot{\w} = 0 \,\, \Rightarrow \,\, \w \mathcal{M}^{-1} \p = 0
            \label{app_eq:constraint_p}
        \end{equation}
        where $\dot{\w}$ stands for the time derivative of the vector $\w$, which we have rewritten in terms of the corresponding momenta $\p = \mathcal{M}\dot{\w}$, with $\mathcal{M}$ being a fictitious diagonal mass matrix. The interpretation is obvious, that is the momenta must be perpendicular at all times to the hypersphere norm vector.
        
        During the numerical integration of the stochastic equations of motion, the unconstrained pseudo--Langevin update generally produces configurations that do not exactly satisfy this condition. To restore the constraint, the configuration must be projected back onto the constraint hypersphere of radius $\sqrt{Q}$. 
        Let's call $\left( \w^*(t + \delta t),  \p^*(t + \delta t)\right)$ the unconstrained update of the current vectors $\left( \w(t), \p(t) \right)$ through pL, with $\delta t$ being the integration time step. The integration scheme is then modified as follows,
        \begin{equation}
            \begin{split}
                \w(t+\delta t) &= \w^*(t + \delta t) + \lambda\frac{\delta t^2}{2} \mathcal{M}^{-1}\nabla C\left( \w(t) \right) \\
                \p(t+\delta t) &= \p^*(t + \delta t) + \\
                               &+ \frac{c_1\delta t}{2} \left[ \lambda\nabla C\left( \w(t) \right) + \Lambda\nabla C\left( \w(t+\delta t) \right) \right]
            \end{split}
        \end{equation}
        where $c_1 $ accounts for the dissipation caused by the viscosity of the medium, $\nabla C(\w) = 2\w$, and $\lambda$, $\Lambda$ are two Lagrange multipliers which can be computed by imposing the validity of Eq. \eqref{app_eq:constraint_w} and Eq. \eqref{app_eq:constraint_p} for $\w(t+\delta t)$ and $\p(t+\delta t)$ respectively.

        First we fix the norm of the updated weight vector by solving
        \begin{equation}
            |\w(t+\delta t)|^2 - Q = 0.
        \end{equation}
        This leads to
        \begin{equation}
            \lambda^{\pm} = \frac{-S_1 \pm \sqrt{S_1^2 - S_3(S_2-Q)}}{\delta t^2 S_3},
            \label{app_eq:lambda_sol_pm}
        \end{equation}
        where we defined
        \begin{equation}
            \begin{split}
                S_1 &= \w^\ast(t+ \delta t) \mathcal{M}^{-1} \w(t) \\
                S_2 &= |\w^\ast(t+ \delta t)|^2 \\
                S_3 &= |\mathcal{M}^{-1}\w(t)|^2
            \end{split}
        \end{equation}
        Among the two roots from Eq. \eqref{app_eq:lambda_sol_pm}, we select as $\lambda$ the one with the smallest absolute value, since it corresponds to the minimal correction compatible with the constraint.

        Once the weight vector has been projected back to the correct hypersphere, we proceed to orthogonalize the momenta vector by solving
        \begin{equation}
            \w(t+\delta t) \mathcal{M}^{-1} \p(t+\delta t) = 0.
        \end{equation}
        Analogously, the solution to this equation is
        \begin{equation}
            \Lambda = -\frac{S_4}{c_1\delta t\,S_5},
            \label{app_eq:Lambda_sol}
        \end{equation}
        where we introduced
        \begin{equation}
            \begin{split}
                S_4 &= \w(t+ \delta t) \mathcal{M}^{-1} \left[ \p^*(t+ \delta t) + \lambda c_1 \delta t \w(t+\delta t) \right]
                \\
                S_5 &= \w(t+ \delta t) \mathcal{M}^{-1}\w(t+ \delta t).
            \end{split}
        \end{equation}

        Lastly, we should mention that a projection is also applied when the initial momentum is generated. Given an unconstrained initial draw $\p(t_0) \sim \mathcal{N} \left( 0, \mathcal{M}T \right)$, we project it onto the tangent space defined by the starting parameters configuration $\w(t_0)$ by subtracting its normal component
        \begin{equation}
            \p(t_0)
            \leftarrow
            \p(t_0)
            -
            \frac{\w(t_0) \mathcal{M}^{-1} \p(t_0)}{|\w(t_0)|^2}\, \mathcal{M}\w(t_0),
        \end{equation}
        which satisfies Eq. (\ref{app_eq:constraint_p}).

\clearpage

\section{Supporting figures}
\label{app_sect:supp_figures}

    \begin{figure}[h!]
        \centering
        \includegraphics[width=0.3\linewidth]{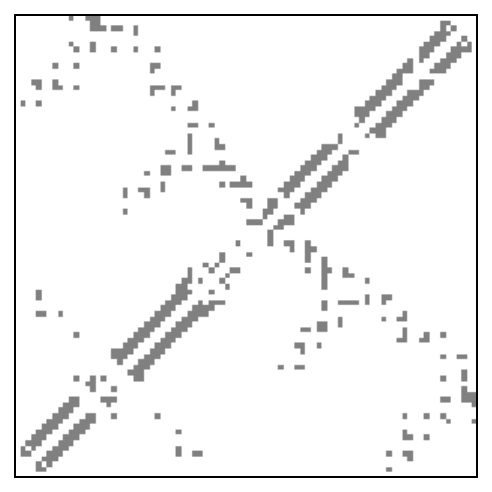}
        \caption{
        The acyl--coenzyme A binding protein contact map.
        }
        \label{app_fig:contact_map}
    \end{figure}
    
    \begin{figure}
        \centering
        \includegraphics[width=0.5\linewidth]{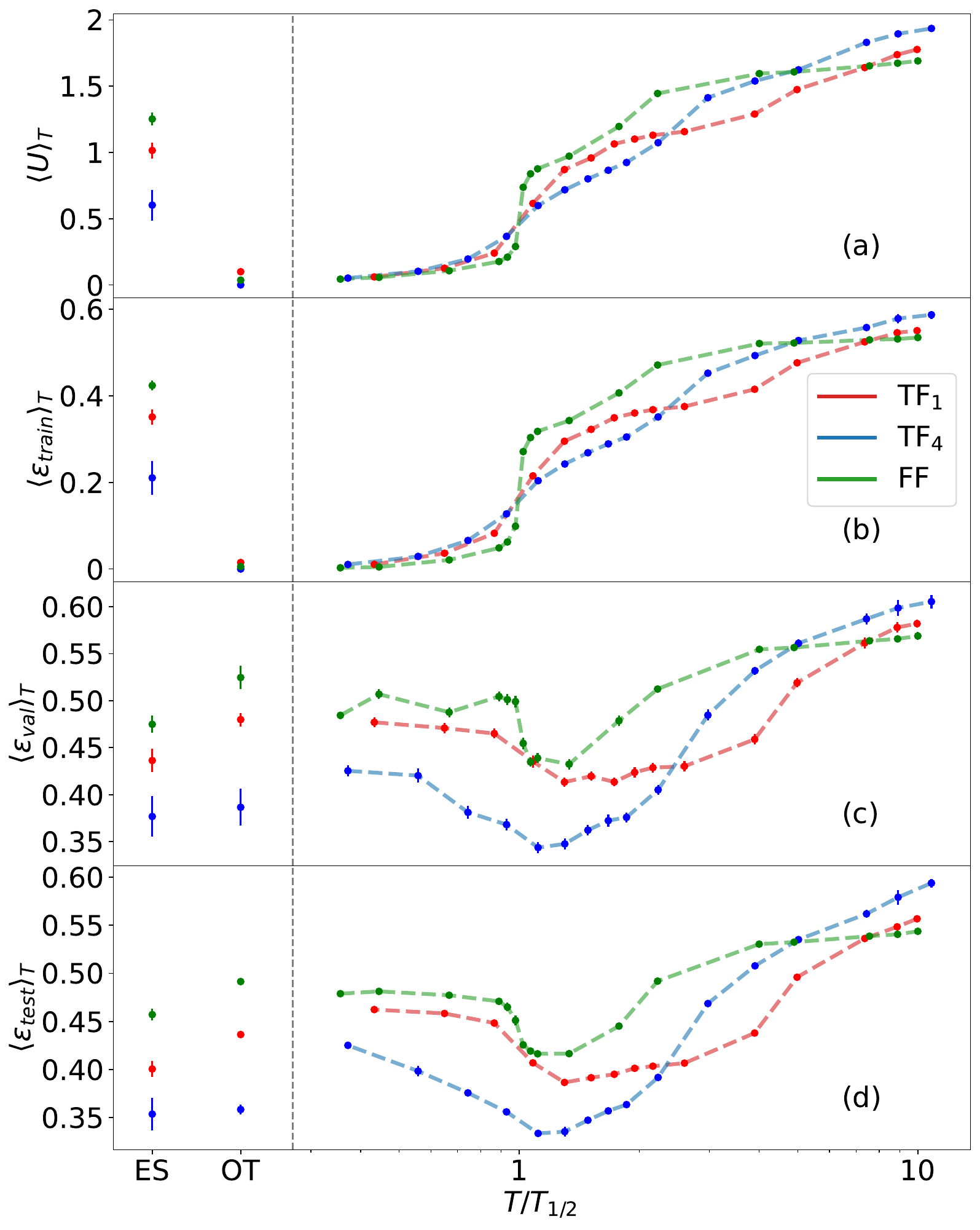}
        \caption{
        Right of the line, the average value of the potential energy $\langle U \rangle_T$ (a), of the training error $\langle \epsilon_{\mathrm{train}} \rangle_T$ (b), of the validation error $\langle \epsilon_{\mathrm{val}} \rangle_T$ (c), and of the test error $\langle \epsilon_{\mathrm{test}} \rangle_T$ (d) as functions of the rescaled temperature $T/T_{1/2}$ for the $\mathrm{TF}_1$ model (red curves), the $\mathrm{TF}_4$ model (blue curves) and the $\mathrm{FF}$ model (green curves).
        Left of the line, the average of the same observables and for the same models, for early–stopped solutions (ES) and over-trained solutions (OT) obtained with Adam.
        }
        \label{app_fig:termodinamica_tot}
    \end{figure}

    \begin{figure}
        \centering
        \includegraphics[width=0.7\linewidth]{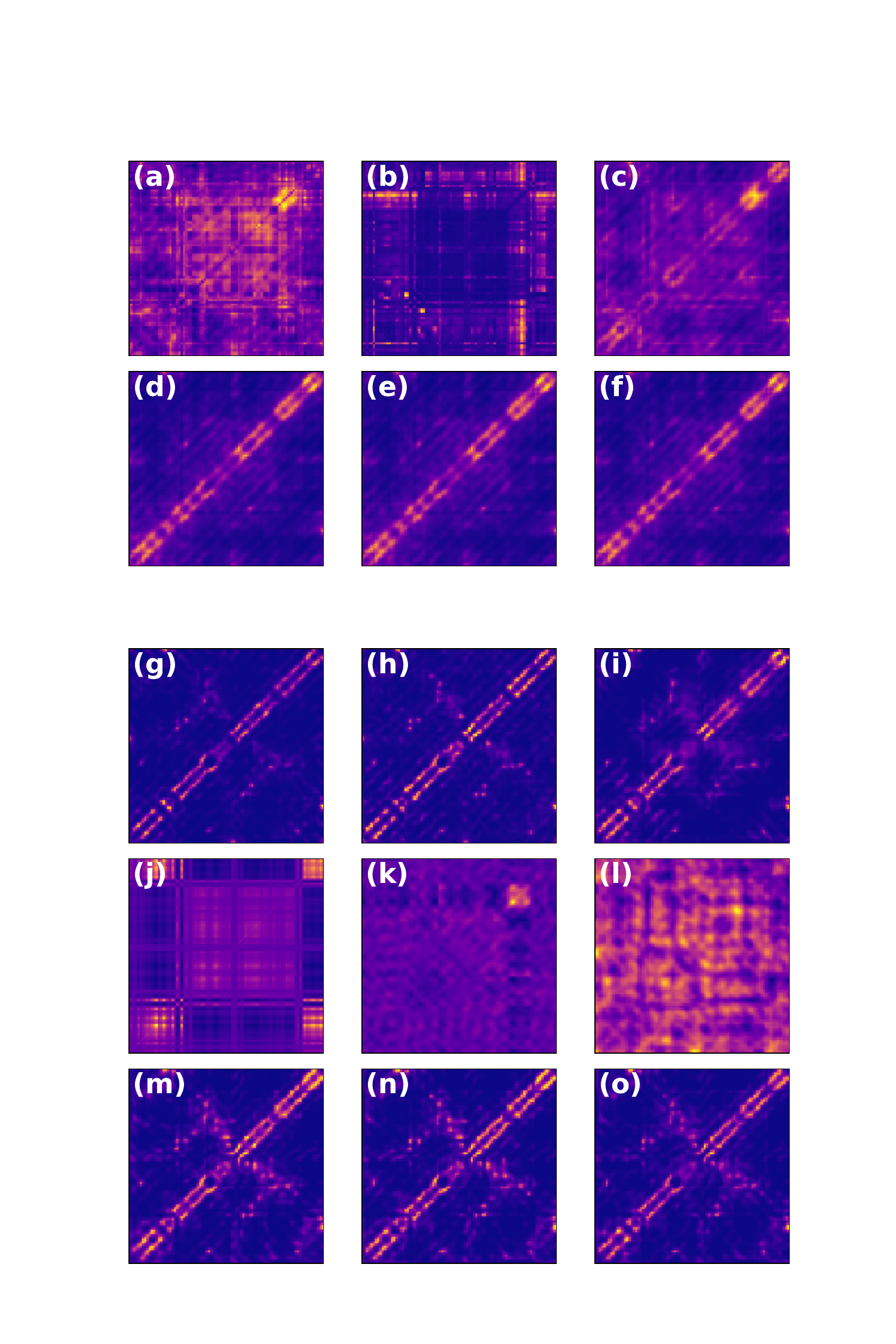}
        \caption{
        Each panel shows a mean attention matrix, that is symmetrized and averaged over the test set sequences.
        The top six images are relative to the $\mathrm{TF}_{1}$ architecture.
        Images (a), (b) and (c) are the mean attention matrices for an early--stopped vector, an over--trained vector, and a single model sampled at $T^\mathrm{best}$.
        Images (d), (e), (f) are the mean attention matrices for three different models sampled at $T=2\,T_{1/2}$.
        The remaining images are relative to the $\mathrm{TF}_{4}$ architecture.
        Images (g), (h), (i) are the best--block mean attention matrices for an early--stopped vector, an over--trained vector and a single model sampled at $T^\mathrm{best}$, respectively.
        Images (j), (k), (l) are the mean attention matrices of the first, second and fourth blocks of the $\mathrm{TF}_{4}$ model sampled at $T=1.1\,T_{1/2}$.
        Images (m), (n), (o) are the mean attention matrices of the third block of the $\mathrm{TF}_{4}$ model for three different models sampled at $T=1.1\,T_{1/2}$.
        }
        \label{app_fig:attention_app}
    \end{figure}
    
    \begin{figure}
        \centering
        \includegraphics[width=0.7\linewidth]{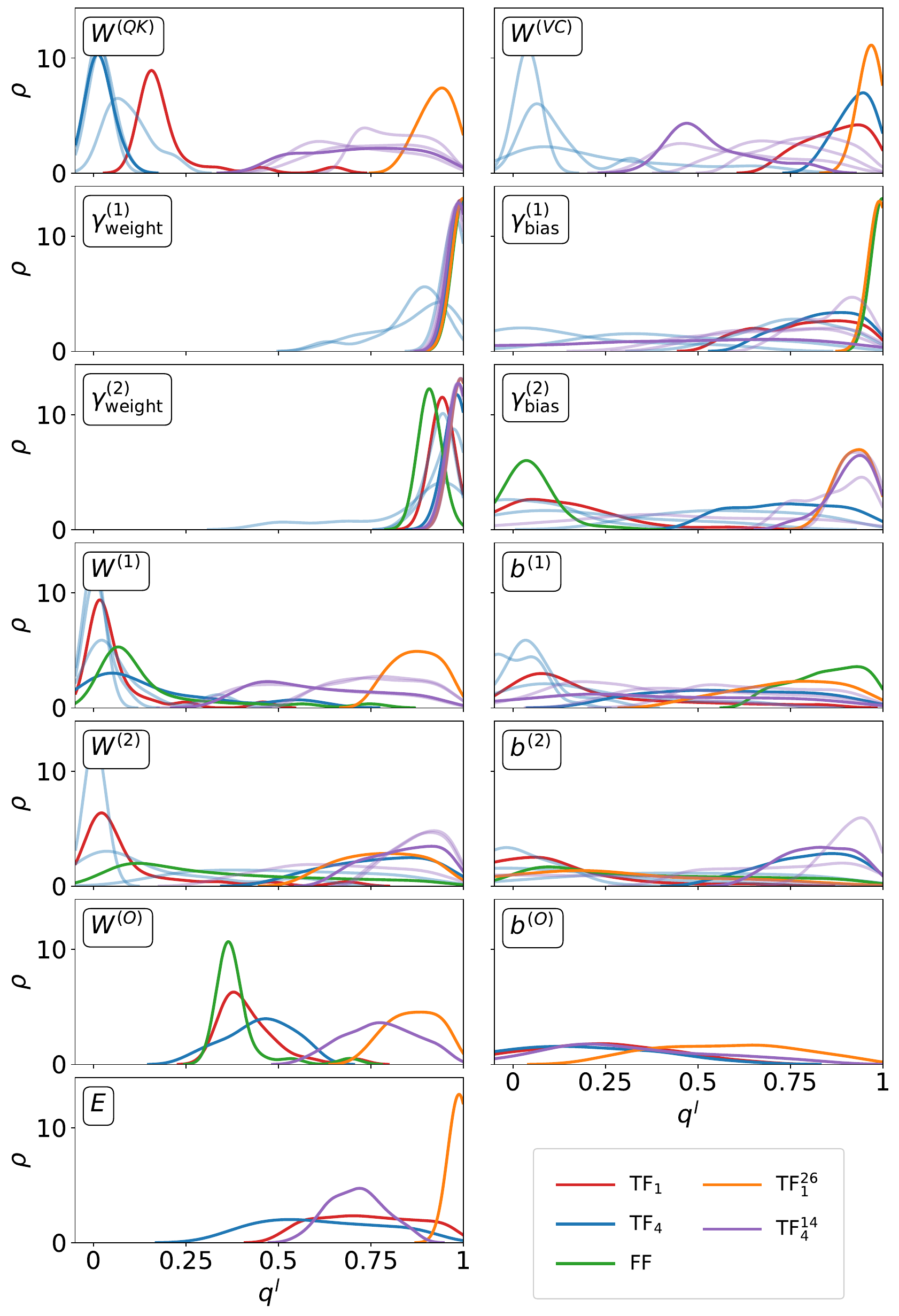}
        \caption{
        The distributions $\rho$ of the similarity $q^l$ calculated between the weights of each layer. The similarity values are obtained by comparing equilibrated configurations at temperature $T^{\mathrm{best}}$, different for each architecture, i.e. $\mathrm{TF}_{1}$ (red curves), $\mathrm{TF}_{1}^{26}$ (orange curves), $\mathrm{TF}_{4}$ (blue curves), $\mathrm{TF}_{4}^{14}$ (purple curves), and $\mathrm{FF}$ (green curves).
        For the transformer-based models with four modules, in the panels for $W^{(QK)}$, $W^{(VC)}$, $\boldsymbol{\gamma}^{(1)}_{\mathrm{weight}}$, $\boldsymbol{\gamma}^{(1)}_{\mathrm{bias}}$, $\boldsymbol{\gamma}^{(2)}_{\mathrm{weight}}$, $\boldsymbol{\gamma}^{(2)}_{\mathrm{bias}}$, $W^{(1)}$, $\boldsymbol{b}^{(1)}$, $W^{(2)}$ and $\boldsymbol{b}^{(2)}$, the solid lines represent the distribution computed between weights from the first block, whereas the three faded curves are relative to the weights from the second, third and fourth block.
        }
        \label{app_fig:q_tot}
    \end{figure}

    \begin{figure}
        \centering
        \includegraphics[width=0.7\linewidth]{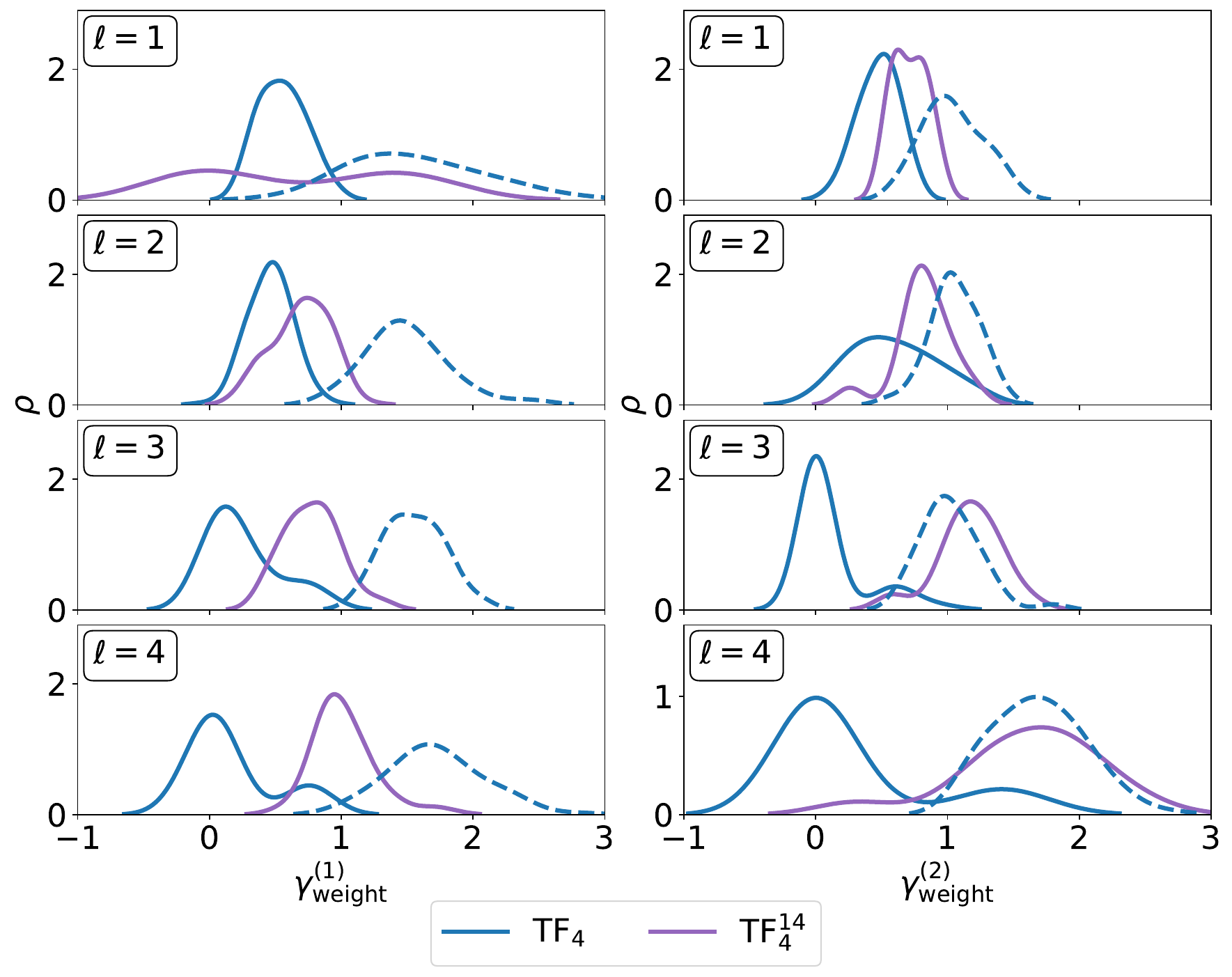}
        \caption{
        Distribution $\rho$ of the values of $\boldsymbol{\gamma}_{\mathrm{weight}}^{(1)}$ (left panels) and $\boldsymbol{\gamma}_{\mathrm{weight}}^{(2)}$ (right panels) for the four--block architectures $\mathrm{TF}_4$ (blue curves) and $\mathrm{TF}_4^{14}$ (purple curves) in the different layers ($l=1, 2, 3, 4$).
        The solid lines represent the results obtained through equilibrated simulations at $T^{\mathrm{best}}$ separately for each architecture.
        The dashed red lines represent the distribution of $\boldsymbol{\gamma}_{\mathrm{weight}}^{(1/2)}$ values obtained from the ES solutions for the $\mathrm{TF}_4$ network only.
        }
        \label{app_fig:gamma_layer}
    \end{figure}
    
    \begin{figure}
        \centering
        \includegraphics[width=0.5\linewidth]{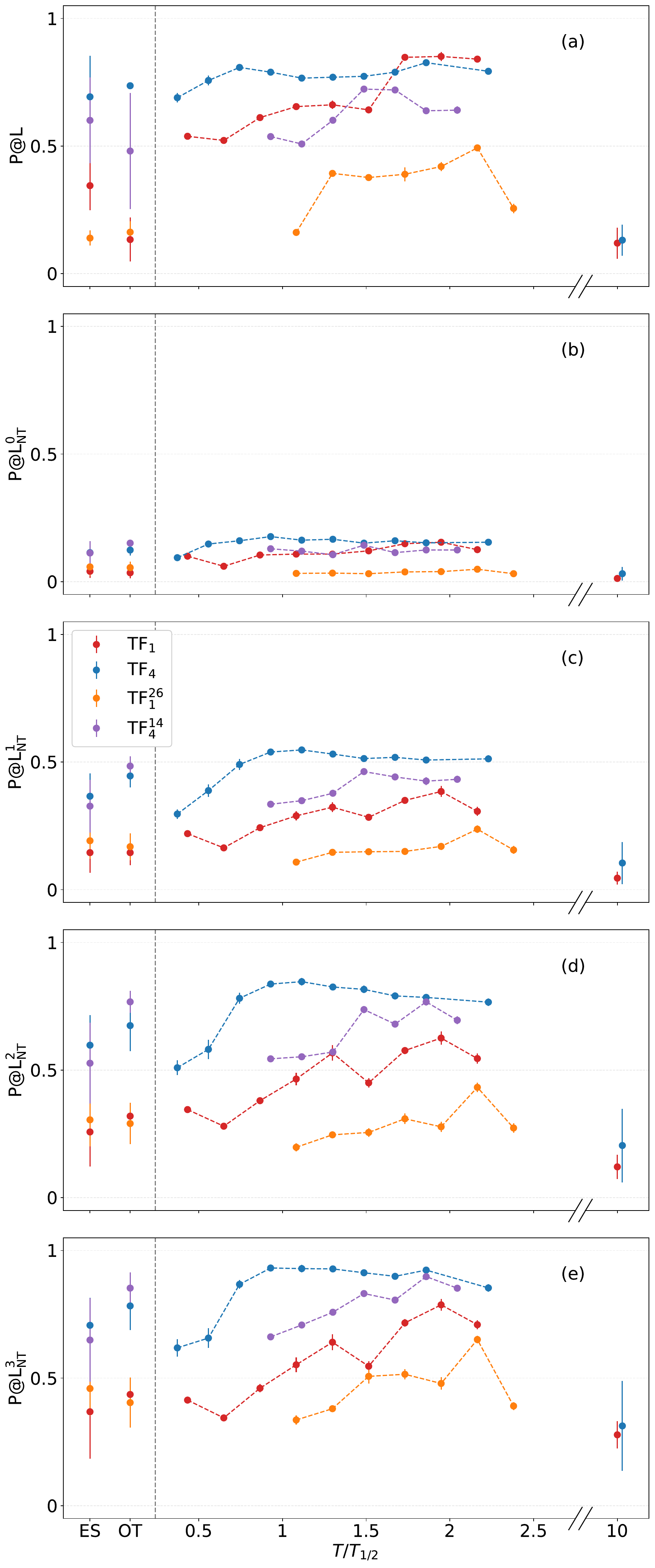}
        \caption{
        The average and standard deviation of the $\mathrm{P@L}$ (a) and of the $\mathrm{P@L}_{\mathrm{NT}}^{\Delta}$ metrics for $\Delta = 0, 1, 2, 3$ ((b), (c), (d) and (e) respectively) for ES and OT vectors (left of the line) and for models obtained from sampling at different rescaled temperatures $T/T_{1/2}$, for each architecture type.       
        }
        \label{app_fig:P@L_tot}
    \end{figure}
    
    \begin{figure}
        \centering
        \includegraphics[width=0.5\linewidth]{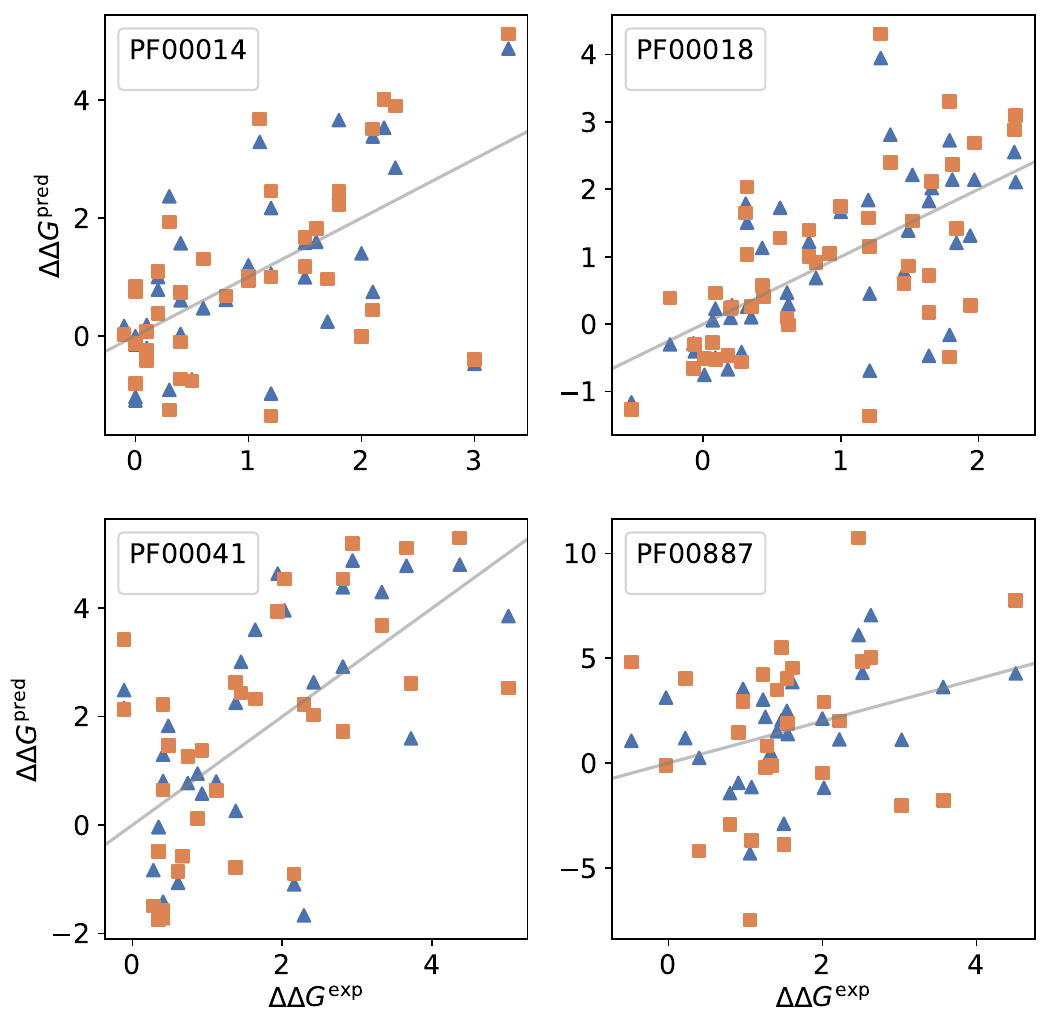}
        \caption{
        Scatterplot of the experimental (x-axis) against the predicted (y-axis) $\Delta \Delta G$ for four protein sequences, each associated to a different protein family (PF00014, PF00018, PF00041 and PF00887).
        In each panel, each point represents a single-site mutation and the predicted free-energy variation $\Delta \Delta G^\mathrm{pred}$ value is obtained using the early-stopped solution (orange squares) or the model sampled at $T^\mathrm{best}$ (blue triangles) with the best Spearman correlation coefficient (see Fig. \ref{fig:Pfam_ddGs_corr}) for that specific sequence.       
        }
        \label{app_fig:Pfam_ddGs_scatterplot}
    \end{figure}


\end{document}